\documentclass[preprint]{aastex} 

\usepackage{graphicx}        
\usepackage{color}           
\usepackage[left]{lineno}

\usepackage{txfonts}
\usepackage{natbib} 
\usepackage{fullpage}
\bibpunct{(}{)}{;}{a}{}{,} 
\begin{document}


\title{On the variation of the scaling exponent of the flare fluence with temperature}

\author{M. Kretzschmar}
       

   \affil{ LPC2E, UMR 7328 CNRS and University of Orl\'eans,
              3a av. de la recherche scientifique, 45071 Orl\'eans, France}
              
                     \email{matthieu.kretzschmar@cnrs-orleans.fr}

\keywords{Flares; Solar Irradiance; Heating, in Flares}

\begin{abstract}
Solar flares result in an increase of the solar irradiance at all wavelengths. While the distribution of the flare fluence observed in coronal emission has been widely studied and found to scale as $f(E)\sim E^{-\alpha}$, with $\alpha$ slightly below 2, the distribution of the flare fluence in chromospheric lines is poorly known. We used the solar irradiance measurements observed by the SDO/EVE instrument at a 10s-cadence to investigate if there is a dependency of the scaling exponent on the formation region of the lines (or temperature). We analyzed all flares above the C1 level since the start of the EVE observation (May 2010) to determine the flare fluence distribution in 16 lines covering a large range of temperature, several of which were not studied before. Our results show a small downward trend with the temperature of the scaling exponent of the PDF, going from above 2 at lower temperature (a few 10$^4$K) to $\sim$1.8 for hot coronal emission (several 10$^6$ K). However, because colder lines also have smaller contrast, we could not exclude that this behavior is caused by including more noise for smaller flare for these lines. We discuss the method and its limits and tentatively associate this possible trend to the different mechanisms responsible for the heating of the chromosphere and corona during flares.
\end{abstract}

\section{Introduction}
     \label{S-Introduction} 
Solar flares are huge release of energy in the atmosphere of the Sun that are frequently observed through bursts of electromagnetic radiation all over the spectrum \citep{Hudson:2011bh,Fletcher:2011ys}. Because flares affect large parts of the solar atmosphere with various phenomena, the emission is coming from plasma in different states and therefore arises at different wavelengths with different contrasts. Two phenomena are usually distinguished during solar flares: 1) the deposition of energy by accelerated particles in the chromosphere and 2) the chromospheric evaporation that leads to strong emission of coronal loops in the extreme-ultraviolet (EUV) and soft X-rays (SXR) wavelength range. The EUV and SXR coronal emission is composed of spectral lines and free--free continuum emitted by a very hot plasma with highly ionized elements. Chromospheric emission arises at different wavelengths. The most contrasted flare signal in the chromosphere are probably the hard X-rays (HXR) (see \textit{e.g.} \citet{Veronig:2002zp}), and is caused by the braking of accelerated electrons in the dense chromosphere. Emission with large contrast and coming from the chromosphere are also observed at radio frequencies, but most of the radiative energy coming from the chromosphere is emitted at shorter wavelengths, in the ultraviolet (UV) and even extreme-ultraviolet wavelength range. Part of the energy initially received by the chromosphere early-on during the flare is quickly used to form bright coronal loops filled with hot plasma, but another part is also radiated at these wavelengths by the chromospheric plasma itself during the impulsive phase. Furthermore, the chromosphere also receives energy by conduction, as well as radiation, from the surrounding coronal loops (\textit{e.g.}, \cite{Berlicki:2004fy,Longcope:2014th}). All these energy transfers bring the chromosphere into a very complex state, and the modeling of the chromospheric emission in the visible and UV during flare remains a very difficult task  \citep{2005ApJ...630..573A,Heinzel:2012kx,2007ASPC..368..387B}. During flares, the whole atmosphere is evolving very rapidly and multiple interactions take place between plasma and photons, which makes often difficult to understand what is going on in the observations (see \textit{e.g.} \citet{Veronig:2010uq} for \textit{e.g.} one -relatively- understandable case and another, more complex case). In this study, we concentrate on the flare emission by chromospheric EUV lines and sometimes refer to it simply as chromospheric emission.\\

 
 Chromospheric evaporation causes the filling of coronal loops by very hot plasma. This leads to strong contrasts in the SXR and short EUV wavelength ranges, even when the light is integrated over the solar disc, \textit{i.e.} in solar irradiance measurements. These SXR and EUV flare signatures are therefore easier to observe than the less contrasted chromospheric emission. When observed at these coronal wavelengths, the flare fluence is well known to be distributed according to a power-law, 
\begin{equation}f(E)\sim E^{-\alpha},\end{equation}
 with $E$ the fluence, $f(E)$ is the probability distribution function (PDF) of the flare fluence, and $\alpha$ the scaling exponent that is usually observed to be slightly below 2 \citep{Hudson:1991aa,Crosby:1993yk,Hannah:2011vn}. For example, \citet{Veronig:2002qf} made an in-depth analysis of Soft X-ray flares observed by GOES and found a power-law frequency distribution for SXR flare fluences with an exponent of about 1.8$\pm$0.1 (background subtracted). In this study, we find for a different period of time a value of 1.86$\pm$0.02 (see Figure \ref{Fig_goes_pdf}). Although small flares are more difficult to observe and measure, there are  indications that small(nano) and large flares belong to the same distribution (\textit{e.g.}, \cite{Hannah:2011vn,Schrijver:2012rt}). HXR emission is produced in the chromosphere but with different processes than EUV lines. The statistics of flares emission in the HXR has also been largely studied and is reviewed by \cite{Hannah:2011vn}. The peak flux values \citep{Dennis:1985kq,Crosby:1993yk} is found to be distributed according to a power law with scaling exponent between 1.5 and 2. \\
 
The value of the scaling exponent $\alpha$ is important since only a value $\alpha > 2$ allows a significant contribution of the smallest flares to the total energy released by all flares \citep{Hudson:1991aa}; nanoflares have been invoked as one way to heat the solar corona, but their ubiquitous nature gives them also the potential to contribute to the variability of the solar radiative flux received at Earth, \textit{i.e.} the solar spectral irradiance. This latest scenario requires however a mechanism to make the (nano) flaring rate vary, which has not yet been observed (but cannot be ruled out because of the in-phase variability of most of the solar activity manifestations with the sunspot cycle). \\
The potential dependency of the scaling exponent $\alpha$ on wavelengths is also important as it would give a constraint on flare models and since it reflects the modification of the solar spectrum shape with the flare size. Indeed, if the flare fluences $E_{\lambda_{1}}$ and $E_{\lambda_{2}}$ measured at two wavelengths $\lambda_1$ and $\lambda_2$ have different scaling exponents, the energy radiated at these two wavelengths will not change at the same rate with the flare size (we detail this in section \ref{S-Scaling}). This parameter $\alpha$ is therefore also important for investigating the effect of flares on the Earth's upper atmosphere, whose the response is also dependent on wavelengths.\\

Because of the smaller contrast and the difficulty to observe the UV chromosphere with space-based instrument (many UV chromospheric lines can be measured from the ground), previous studies of the distribution of the flare energy have focused on observations at coronal wavelengths or in the HXR and radio domains, which have different physical mechanisms than the chromospheric lines. There has been no study to our knowledge dealing with the statistics of flare fluences in E/UV chromospheric lines. Measuring the flare fluence for an ensemble of flares requires a certain level of calibration that is difficult to obtain from the ground. \textit{E.g.} \citet{Temmer:2001kq} presented a statistical analysis of H$\alpha$ flare, characterizing various parameters, but without reporting on the flare fluence for this line. However, chromospheric emission can represent a large part of the total energy radiated by flare \citep{Woods:2006aa, Kretzschmar:2010lr, Kretzschmar:2011lr, Fletcher:2011ys, Milligan:2012fj, Milligan:2014mz} and determining $\alpha$ for these emissions is therefore important. \\

The main aim of this paper is to investigate the scaling of the flare fluence at different wavelengths, searching in particular for potential differences for EUV spectral lines that are predominantly from coronal or chromospheric origin. Recent observations by the \textit{Extreme Ultraviolet Variability Experiment} (EVE; \citet{Woods:2010zr}) onboard the \textit{Solar Dynamic Observatory} (SDO) gives us an excellent opportunity to investigate the flare distribution in EUV chromospheric lines. SDO/EVE is observing the Sun quasi-permanently since May 2010, providing high-resolution irradiance spectrum with a 10 s cadence and at EUV wavelengths representing both the (hot) corona and the (warm) upper chromosphere and lower transition region. In this paper, we use four years of flare observations by SDO/EVE to investigate if and how the flare fluence depends on wavelength and temperature.  
 
 \begin{figure}    
   \centerline{\includegraphics[width=0.8\textwidth,clip=]{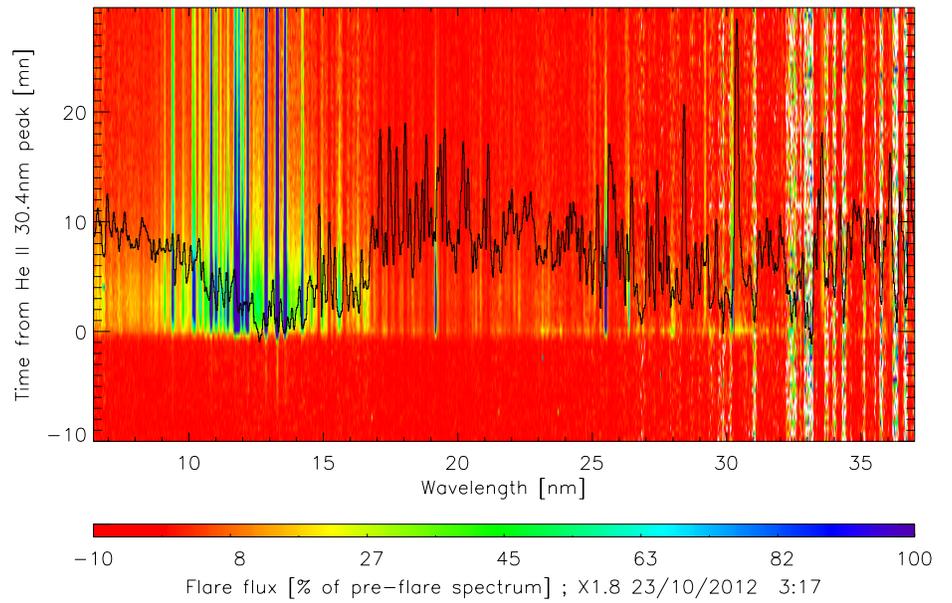} }   
   \centerline{\includegraphics[width=0.8\textwidth,clip=]{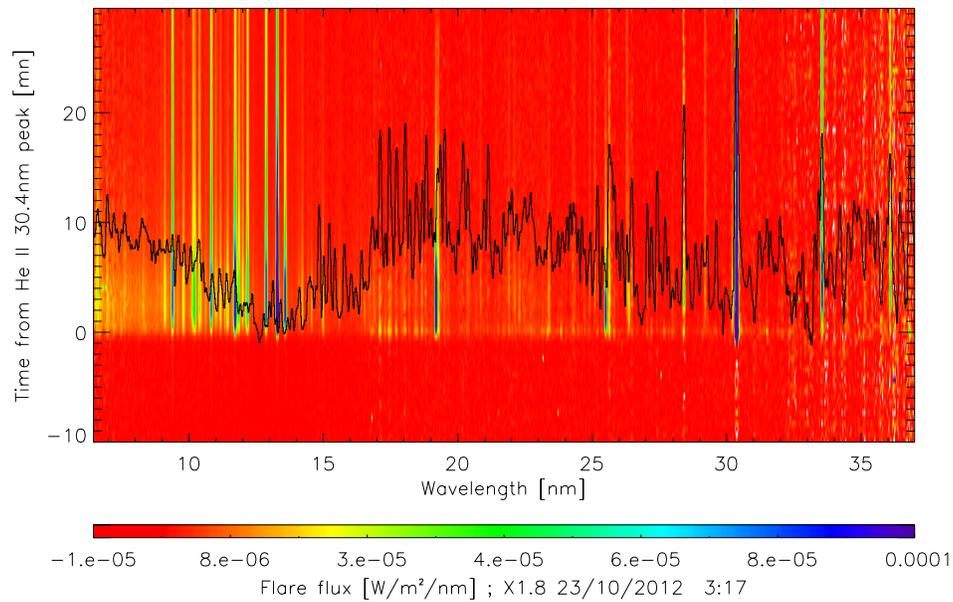} }
              \caption{ Flare flux observed by SDO/EVE for the X1.8 flare of the 23 October 2012, in function of time and wavelength. Time is given in minutes from the peak of the flare in the He II 30.4 nm line. The black line shows the spectrum (no units) for reference. The color represents the intensity in the following way. Top panel : The flare flux is expressed in  \% of the pre-flare flux, which is equivalent to the irradiance contrast of the flare. Bottom panel : The flare flux is expressed in physical units W/m$^{2}$/nm. }
   \label{Fig_flare_ex}
   \end{figure}

 \begin{figure}    
   \centerline{\includegraphics[width=0.9\textwidth,clip=]{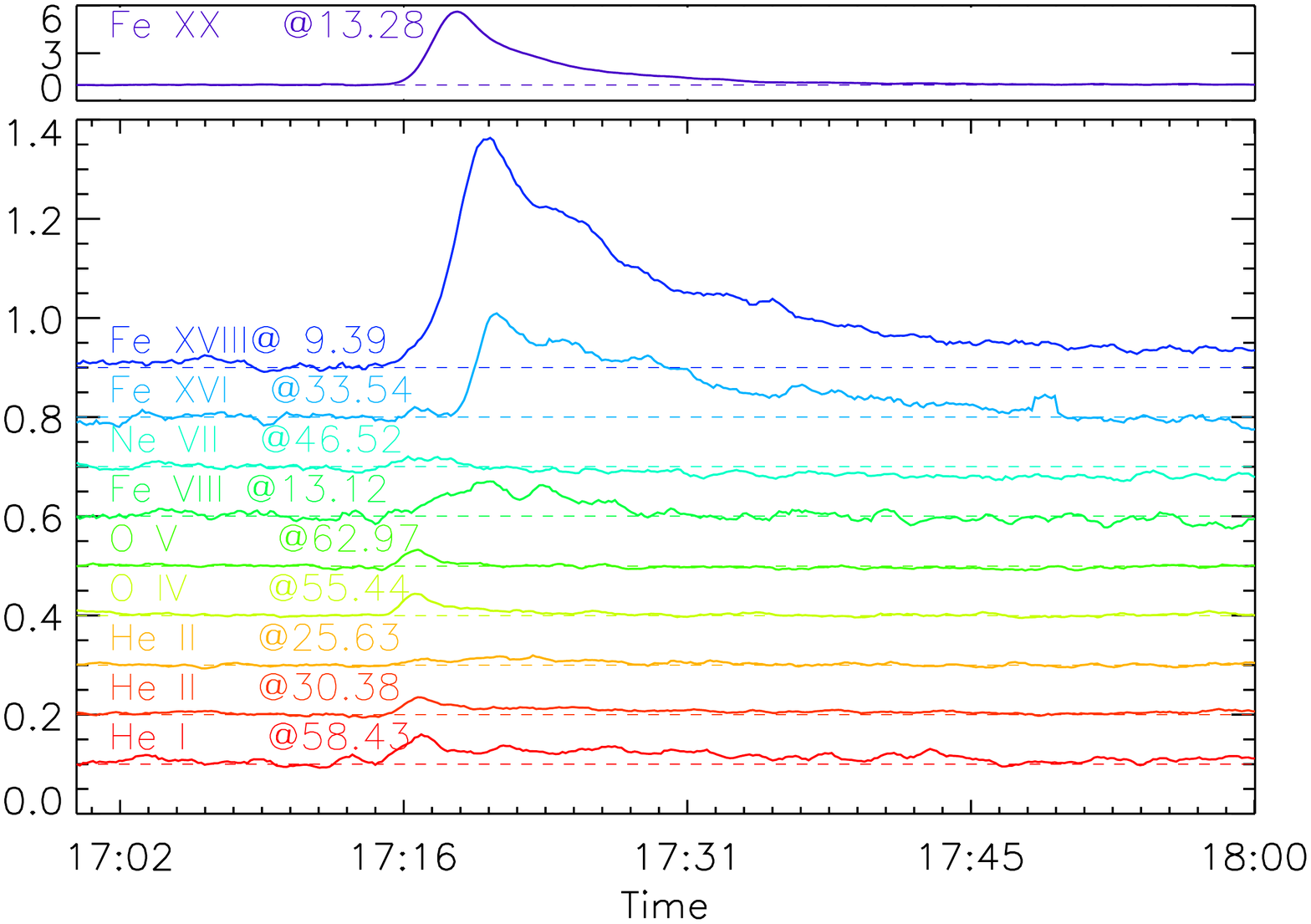} }   
               \caption{ Lightcurves of relative increase (or irradiance contrast) for different lines for the M1.2 flare of 5 May 2010. The Fe XX line at 13.28 nm has been isolated in the upper panel because of its large contrast. In the lower panel, all curves have been shifted vertically by 0.1 for clarity but an increase of 0.1 corresponds to an increase of 10\% .}
   \label{Fig_flare_ex2}
   \end{figure}

 Section~\ref{S-dataanalysis} presents the SDO/EVE data and the spectral lines that are used in the analysis. We also describe the computation of the flare fluence and the determination of the scaling exponent of the fluence distribution.
Section~\ref{S-results} shows and discusses the results while we conclude in Section \ref{S-conclu}.

\section{Data analysis}
 \label{S-dataanalysis}

\subsection{EVE flare spectrum and line fluxes.}

We used the version 4 of the EVE data product, which contains observations from the two spectrometers MEGS-A (5--37 nm) and MEGS-B (35--105 nm) with a spectral resolution of 0.02 nm and from several spectrophotometers (ESP) having broad passbands between 0.1 nm and 39 nm. \\

An example of flare observation by EVE is presented in Figure \ref{Fig_flare_ex} and Figure \ref{Fig_flare_ex2}. Figure \ref{Fig_flare_ex} shows the EVE flare spectrum at the full wavelength resolution in two forms: the upper panel represents the flare (irradiance) contrast in \% while the bottom panel shows the flare flux in physical units. These two representations allow us to realize that the flare energy can reside at wavelengths where the contrast is moderate, as it is for example the case for the Fe XVI line at 36.07 nm. We can also note by looking at the top panel that noise will disturb the flare signal when the contrast becomes smaller than a few \%. Figure \ref{Fig_flare_ex2} shows time series for spectral lines available in the level 2 of the EVE data product; these line fluxes are integrated with fixed spectral integration limits. With no surprise, the hottest lines have the largest contrast. When moving to colder (chromospheric and transition region) lines, \textit{e.g.} the O V line and below in the figure, the flare profile appears to be more impulsive. For this M1-class flare, the contrast of the colder lines is of the order of 5\% in these 1-minute smoothed data. This is comparable to the noise level and for these low-contrast lines only flares above the M1 level can be processed meaningfully through our analysis. Flares were identified using the GOES flare catalog and we first considered the 4664 flares above the C level that were observed by EVE from May 2010 to mid 2014. As explained below, the low contrast of some EUV lines forced us to use only the 435 M- and X-flares during this period.  \\

In the level 2 of EVE data products, the fluxes of several spectral lines are already provided together with the full spectrum. These lines are however integrated with a fixed spectral width on each spectrum. We used these line-flux values but we have also computed the fluxes directly by integrating the lines observed in the spectrum, in order to deal with possible modifications of the spectrum and line shapes during the flare. This was tested in two different ways; first, by determining automatically the spectral integration limits as follow: we started from the peak of the line and defined the blue and red limits of the integration by identifying the first wavelength at which the intensity stops decreasing when moving away from the peak. Second, we attempted to fit the lines with a continuum and a Gaussian profile. This second method gave valuable but also very intermittent results : several spectral lines needed to be fitted at each time step of a flare, and the fit failed several times during this procedure, which make the final statistics difficult to interpret. At the end, we used the line intensities provided in the EVE data product and the ones that we computed by direct integration (with automatic determination of the spectral limits). We required the line-flux values retrieved by these two methods to agree in order to keep them in the statistics.   \\

Several of the lines available in the EVE level 2 data product have very low flare contrast and/or do not show a clear scaling of the flare signal with the SXR class of the flare (see also Figure \ref{Fig_SNR} and Section \ref{S-SNR}). Also, because of the strong degradation of the MEGS-B spectrometer, observations above 35 nm has been reduced to three hours per day, and consequently fewer flares are observed at these wavelengths. We kept only two of these lines (the O IV line at 55.44 nm and the Si XII line at 49.94 nm). Furthermore, because during flares the dominant spectral lines of the spectrum can change with respect to those of a quieter Sun, we analyzed a few flare spectra to determine other lines of interest, with the objective to cover the largest range of temperature possible. These lines were identified with the use of the CHIANTI database \citep{1997A&AS..125..149D,2006ApJS..162..261L}, by simulating two spectra for flare and active region in order to ensure that the dominant line in the integrated spectral bins is the same for large and small flares. The flare spectrum simulated with the CHIANTI database is based on a differential emission measure (DEM) computed from a flaring region of the Sun \citep{Dere:1979fj} and not an integrated Sun; therefore we do not expect to observe the same lines in the EVE irradiance spectrum but we used the ratio of the flare intensity to the active region intensity as a criterium to determine which line is dominant. Table \ref{T-Lines} shows the lines that we finally used, with the lines that are not present in the EVE data product marked with an asterisk. \\

 \begin{table}
\caption{List of spectral lines that we retained as usable for this study. The Asterisk indicates lines that are not already provided in the EVE data product. }
\label{T-Lines}
\begin{tabular}{lcc}     
  \hline                   
Ions & Wavelength (nm) & Log(T) \\
  \hline
He II   & 30.38  &    4.7 \\
O IV   &  55.44   &   5.19 \\
O VI* & 15.01 & 5.5   \\
Fe VIII* & 10.37 & 5.6  \\
Fe VIII & 13.12   &   5.57 \\
Fe IX* & 10.52 & 5.8  \\
Fe XI* & 10.06 & 6.1  \\
Ni XI* & 14.89 & 6.1  \\
Si XII  & 49.94   &   6.29 \\
Fe XV  &  28.42  &    6.3 \\
Fe XVI  & 33.54   &   6.43 \\
Fe XVII* & 25.49 & 6.6\\
 Fe XVIII  & 9.39 &     6.81 \\
Fe XX  &  13.28   &   6.97 \\
Fe XXI*& 12.87  & 7.0 \\
Fe XXII*& 13.57  & 7.1 \\
 \hline
\end{tabular}
\end{table}

\subsection{Determination of the power-law exponent }\label{S-Scaling}
 \begin{figure}    
   \centerline{\includegraphics[width=0.9\textwidth,clip=]{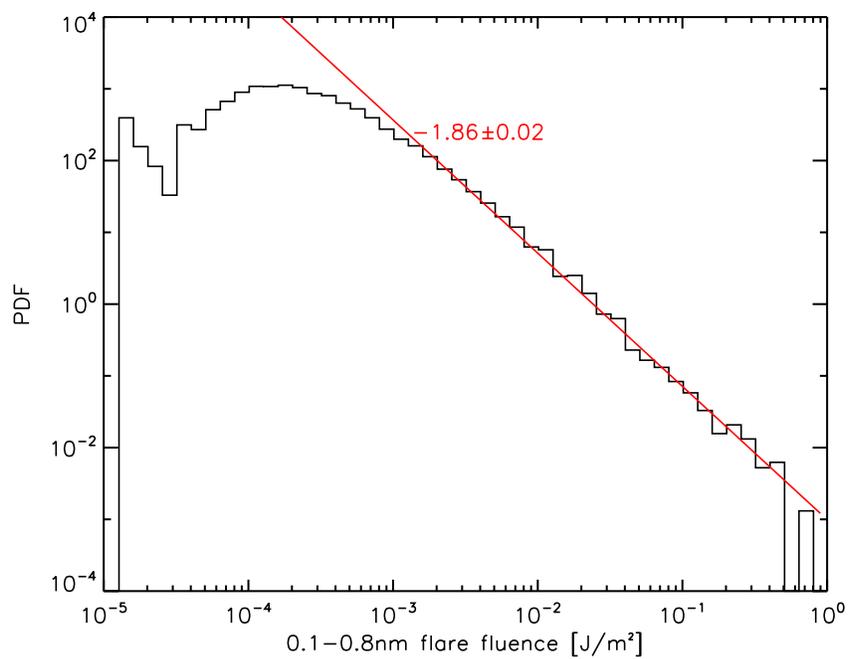} }   
               \caption{ Probability Distribution Function (PDF) of flare fluence in the 0.1-0.8 nm range.}
   \label{Fig_goes_pdf}
   \end{figure}

 \begin{figure}    
   \centerline{\includegraphics[width=0.6\textwidth,clip=]{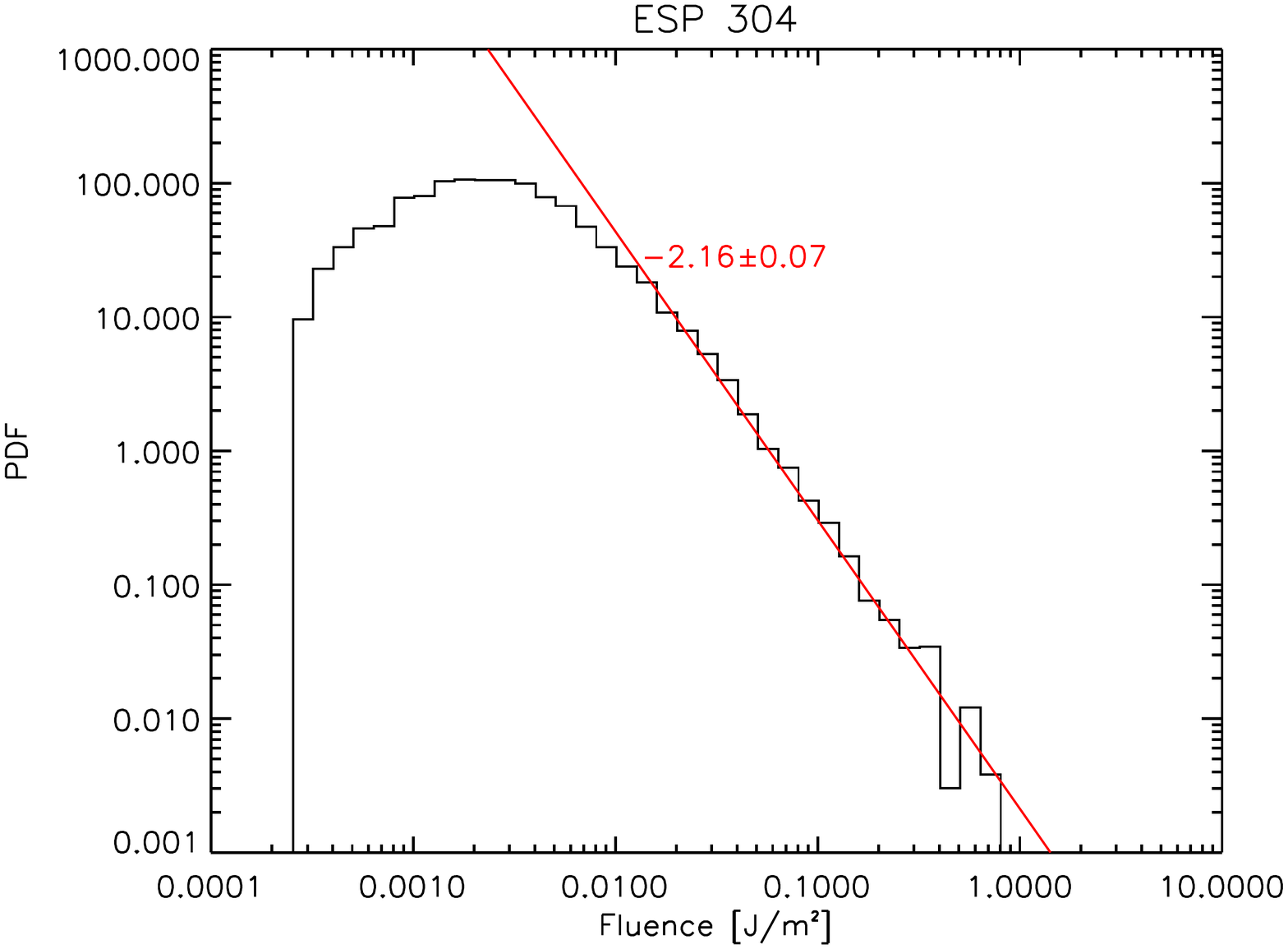}}
  \centerline{\includegraphics[width=0.6\textwidth,clip=]{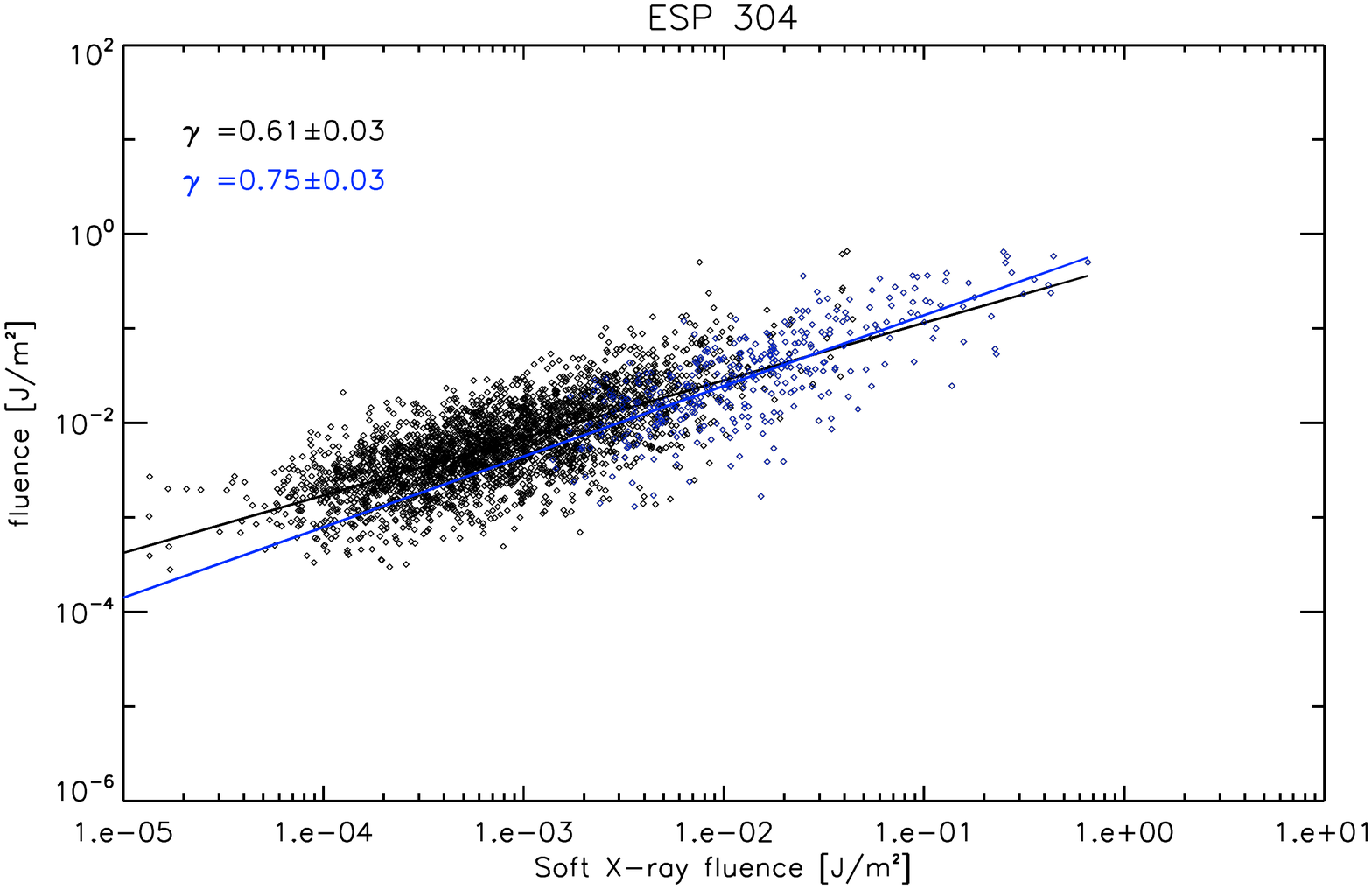}	 }   
            \caption{ Top panel: Probability Distribution Function (PDF) of the flare fluence for the EVE MEGS 304 diode. In red, the best fit to the power-law gives an exponent of 2.16. Bottom panel: Scaling of  EVE MEGS 304 diode flare fluence versus 0.1--0.8 nm. Only flares with a signal-to-noise ratio above two sigma are plotted. Flares above the M1 level are in blue. The two fits are for flares above the C1 and M1 levels respectively in black and blue.
              }
   \label{Fig_304}
   \end{figure}
 \begin{figure}    
   \centerline{\includegraphics[width=0.6\textwidth,clip=]{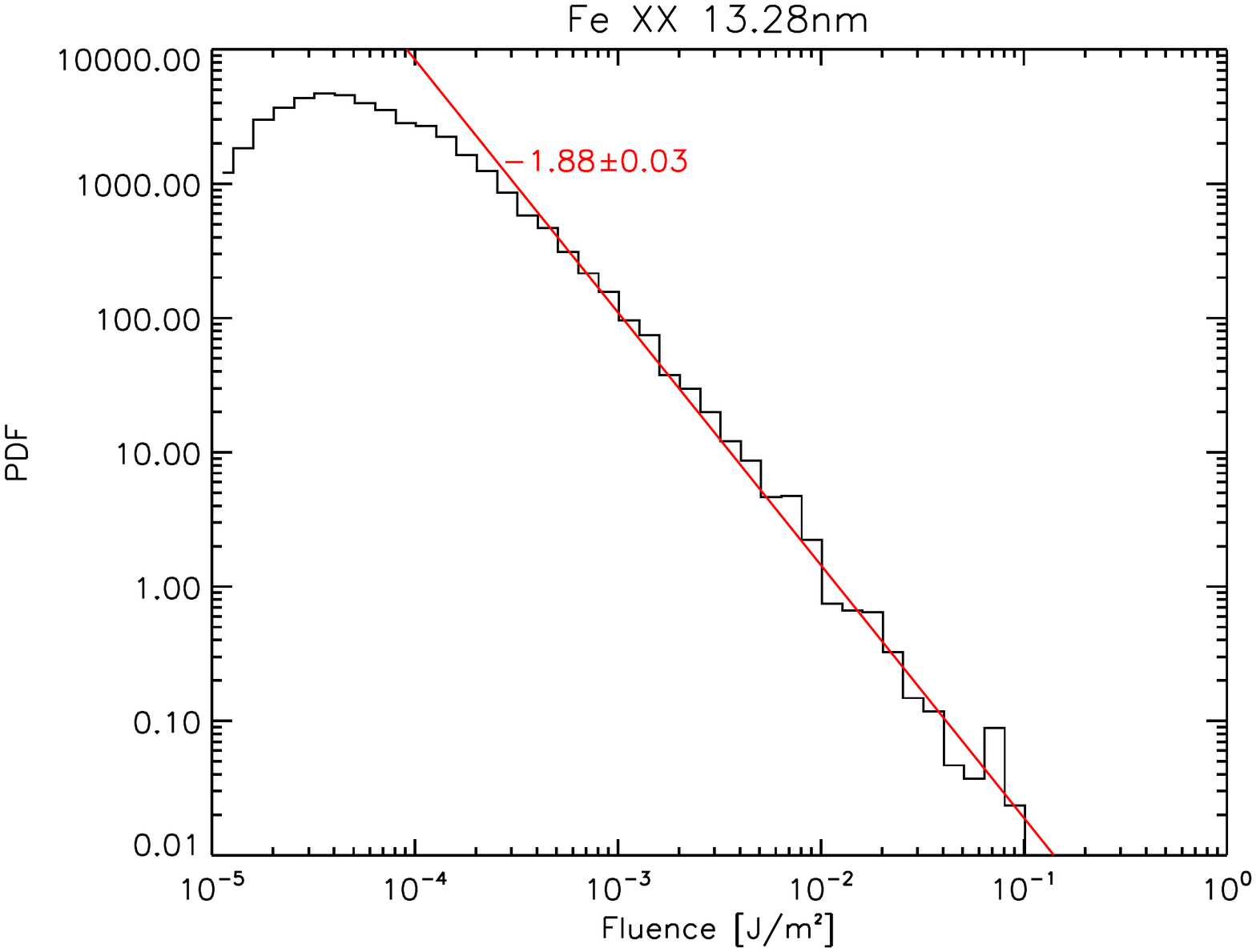}}
  \centerline{\includegraphics[width=0.6\textwidth,clip=]{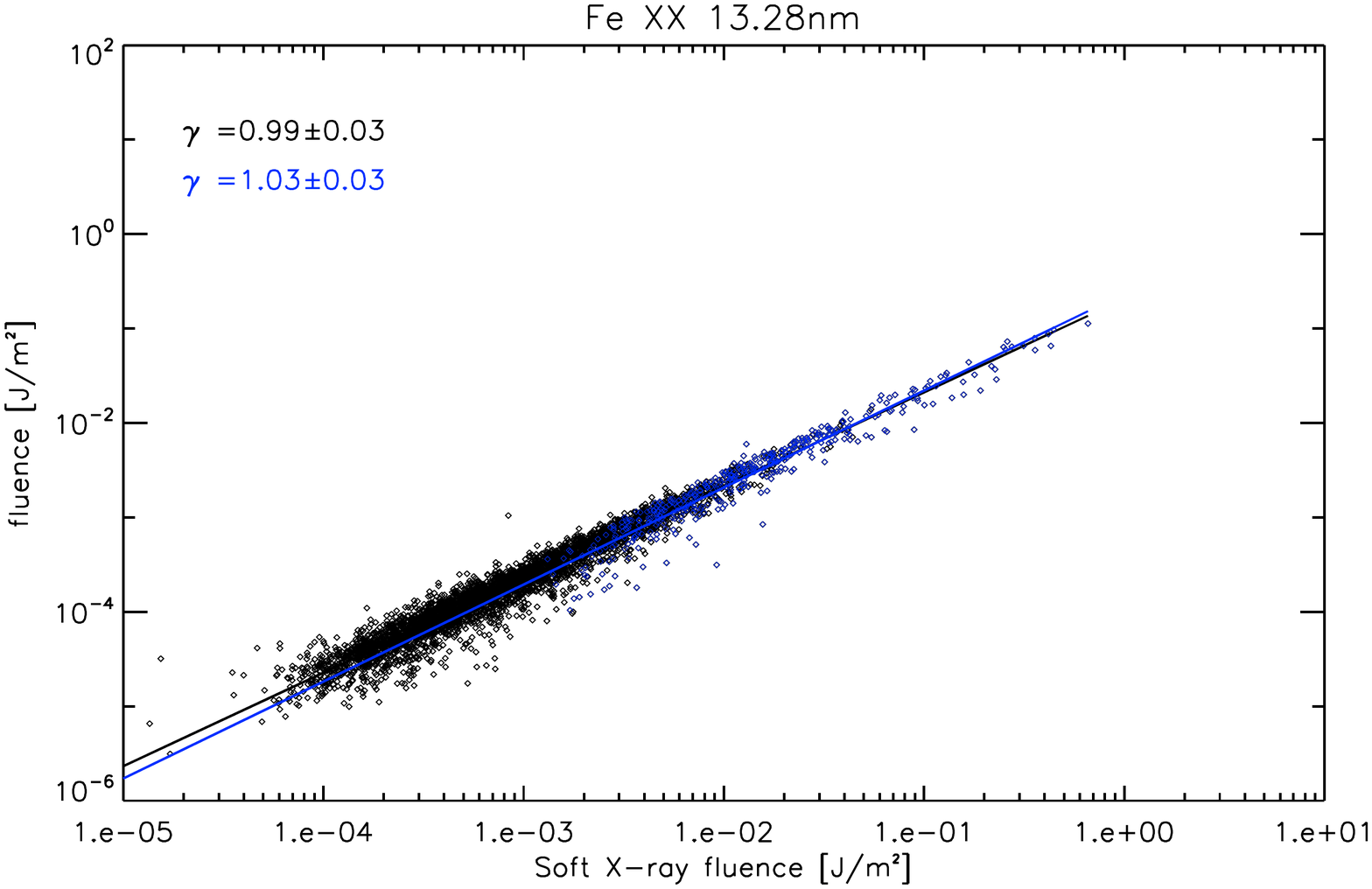}	 }   
               \caption{ Top panel: Probability Distribution Function (PDF) of flare fluence for the EVE Fe XX line at 13.28 nm. In red, the best fit to the power-law gives an exponent of 1.88. Bottom panel: Scaling of the EVE Fe XX line at 13.28 nm flare fluence versus 0.1--0.8 nm flare fluence. Only flares with a signal-to-noise ratio above two sigma are plotted. Flares above the M1 level are in blue. The two fits are for flares above the C1 and M1 levels respectively in black and blue.
               }
   \label{Fig_Fe20}
   \end{figure}

We start this section by making explicit the relationships between the scaling exponents $\alpha_x$ and $\alpha_y$ of two power-law distributed variable $x$ and $y$. Let us first assume that $x$ and $y$ are the flare fluences at two different wavelengths (SXR and EUV \textit{e.g.}), and that: 
\begin{itemize}
\item $x$ is distributed according to a power-law with scaling exponents $\alpha_x$ : $P(x) =C_1 x^{-\alpha_x}$
\item $x$ and $y$ are related through $y \sim x^{\gamma}$. 
\end{itemize}
As $ P(y)dy = P(x) dx$, it is straightforward to verify that 
\begin{equation} \label{Eq-expo} P(y)=C_2 y^{-\alpha_y}$$ with $$\alpha_y=\frac{\alpha_x + \gamma -1}{\gamma} .\end{equation}
 The uncertainty $U(\alpha_y)$ on $\alpha_y$ can be computed with
\begin{equation} \label{Eq-expo-Unc}  U(\alpha_y)= \frac{U(\alpha_x)}{\gamma} +|1-\frac{\alpha_x}{\gamma}-\frac{\gamma-1}{\gamma}| \frac{U(\gamma)}{\gamma} . \end{equation}
We therefore have two ways to estimate the exponent of the distribution function for the flare fluence of one line. The first one is to compute the probability distribution function (PDF) and to fit it with a power-law. The second one, which is particularly relevant when the number of flares with high enough contrast is too small to build a reliable PDF, is to plot the flare fluence of the line versus the flare fluence observed in the 0.1 nm--0.8 nm by GOES, and then apply Equation (\ref{Eq-expo}). Indeed the GOES SXR fluence is well known to have a power-law distribution with exponent slightly below 2 (see \cite{Veronig:2002qf} and reference therein). For the flares that we are considering in this study, we found $\alpha_{GOES}=1.86$ (see Figure \ref{Fig_goes_pdf}), which was determined by using the 1-minute data of the GOES satellite. \\

This procedure is illustrated in Figs.\ref{Fig_goes_pdf}, \ref{Fig_304} and \ref{Fig_Fe20}. The power-law distribution of flare fluence in the 0.1-0.8 nm range has an exponent of 1.86$\pm$ 0.02 while, for the flare fluence observed by the EVE/ESP diode in a wavelength range around the He II 30.4 nm line, we found an exponent of 2.16$\pm$0.07 (top panel of Figure \ref{Fig_304}). By scaling the He II flare fluence to the SXR flare fluence we found the relationship $E_{HeII} \sim E_{SXR}^{\gamma}$ with $\gamma = 0.75\pm0.03$ (bottom panel of Figure \ref{Fig_304}), when considering flares above the M1 level only. By applying the relationship eq.\ref{Eq-expo} between the exponents above, this value of $\gamma$ leads to $\alpha_{304} = 2.15\pm0.07$ for the value of the power-law exponent of the PDF of the flare fluence at He II 30.4 nm. This compares very well to the value of 2.16$\pm$0.07 found by fitting the PDF. Let us also note that $\gamma$ decreases when we include C-class flares, or equivalently that the scaling exponent increases. We attribute this to the fact that for these smaller flares, part of the flares observed at 30.4 nm are below the noise level between the GOES start and end-times; the flare fluence is then artificially increased by the noise.  \\

The same procedure is illustrated for the hot coronal emission produced by the Fe XX ion at 13.28 nm in Figure \ref{Fig_Fe20}. By applying again Equation (\ref{Eq-expo}), the scaling of the Fe XX line fluence to the SXR fluence leads to a value of 1.83$\pm$0.04 for the scaling exponent of the PDF, in good agreement with the value 1.88$\pm$0.03 found by fitting the PDF. Here, the $\alpha$ value found with the C-class flares included is very similar (1.87$\pm$0.04), suggesting that this line has good contrast all over the flare duration as defined by GOES. \\

Computing a robust power-law distribution requires many flares observations but only hot coronal lines have a large contrast enough to compute their flare fluence over several flare magnitudes. Therefore, we first determined $\gamma$ by scaling the flare fluence of a spectral line $E(\lambda)$ to the flare fluence in the SXR $E_{SXR}$ and next applied Equation (\ref{Eq-expo}) to compute the scaling exponent $\alpha$ of the fluence distribution.

\subsection{Flare fluence computation}  \label{S-SNR}
 \begin{figure}    
   \centerline{\includegraphics[width=0.45\textwidth,clip=]{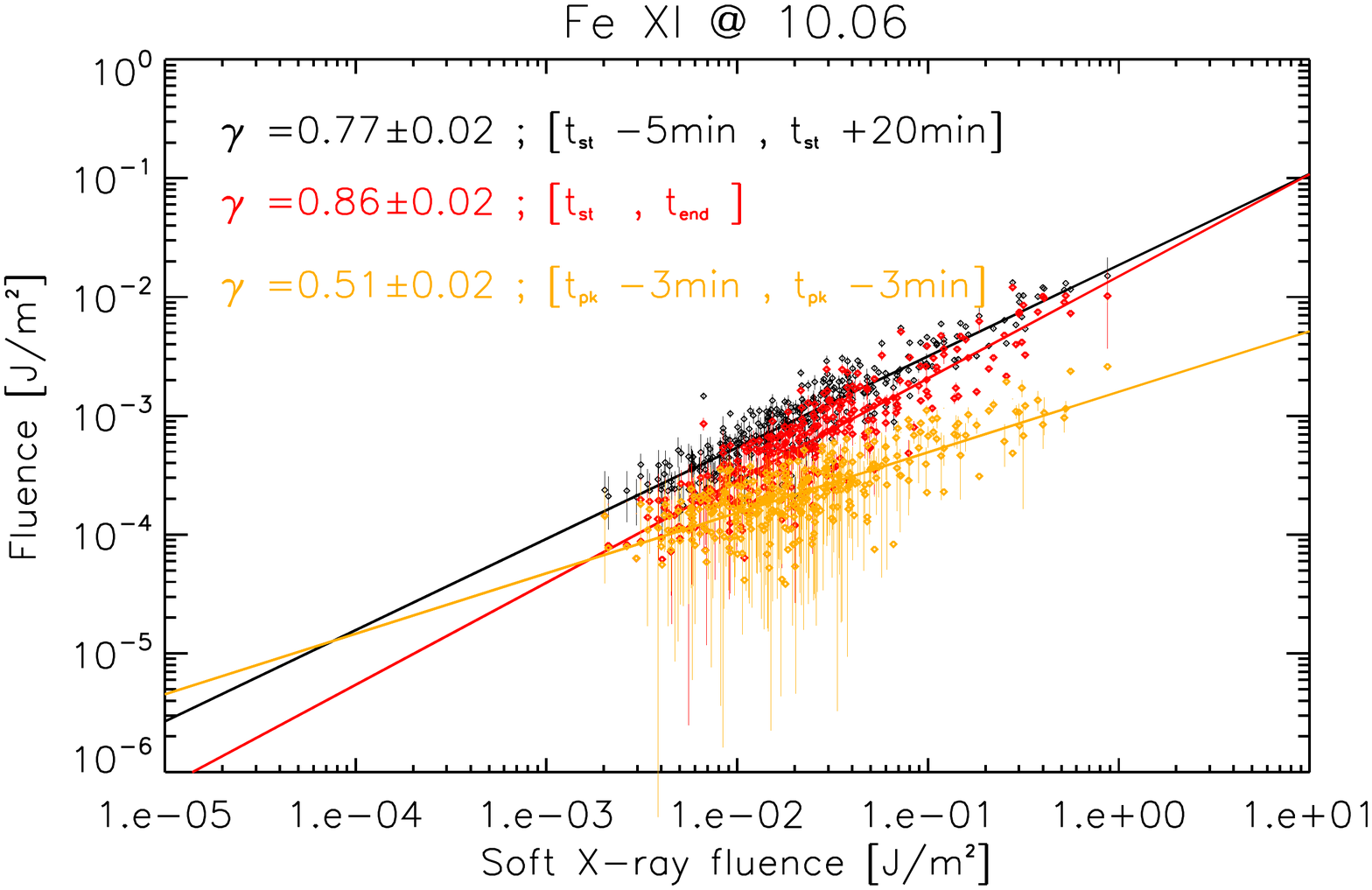} 
   			\includegraphics[width=0.45\textwidth,clip=]{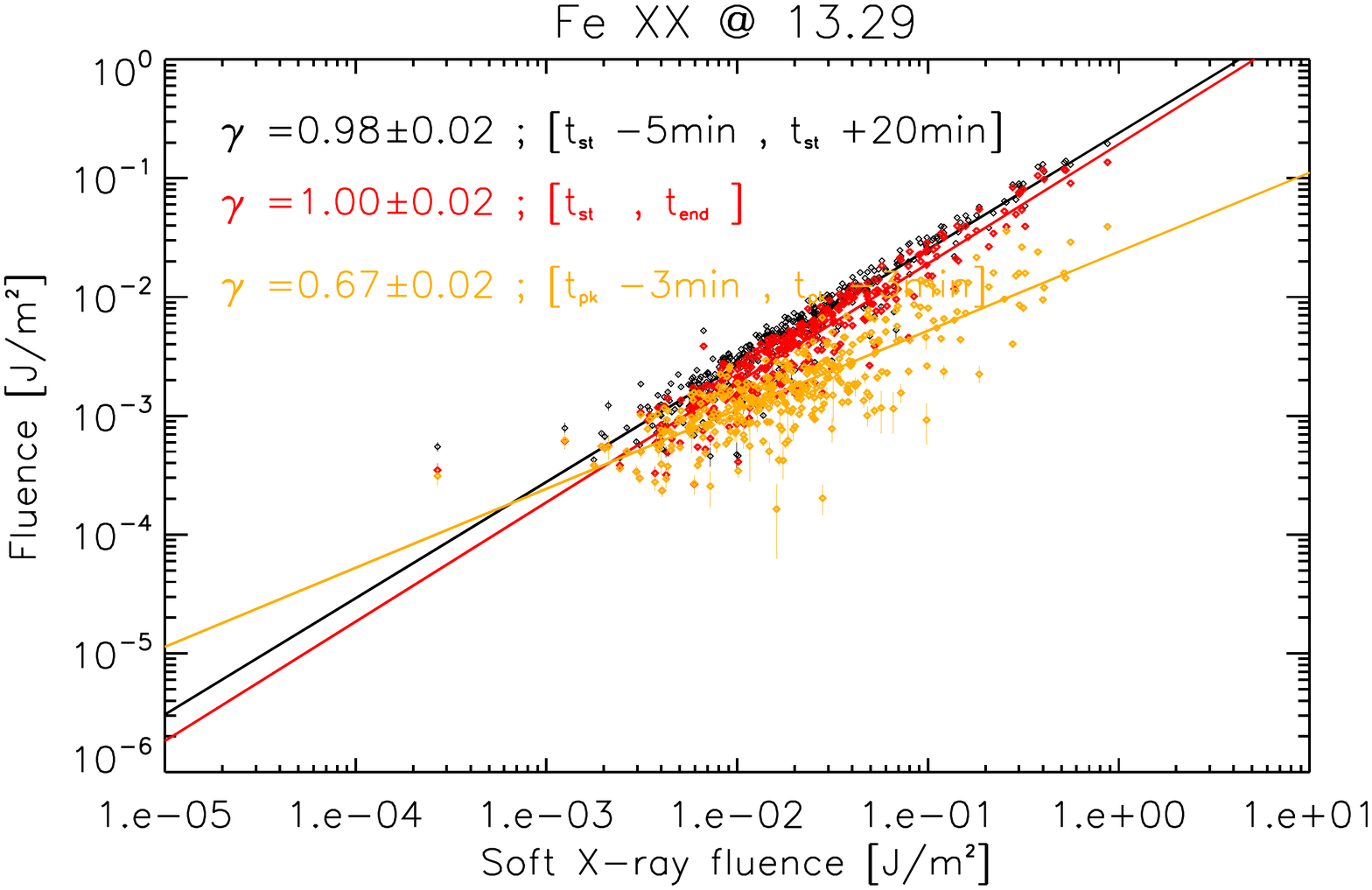} }   
               \caption{Influence of the integration time on the computation of fluence and on the scaling parameter for Fe XI at 10.06 nm (left panel) and Fe XX at 13.28 nm (right panel). Fluences are computed by integration between 1) 5 minutes before the GOES start-time and 10 minutes after the GOES end-time (black diamonds) 2) the GOES start-time and the GOES end-time (red diamonds) 3) 6minutes centered around the GOES peak time}
   \label{Fig_TIME}
   \end{figure}
 
The flare fluence is defined as follows
\begin{equation} \label{Eq-Flu1} E(\lambda)=\int_{t_{st}}^{t_{end}}\left(I(\lambda,t)-I_{bg}(\lambda)\right)dt ,\end{equation}
where $I(\lambda,t)$ is the irradiance at time $t$ during the flare, $I_{bg}(\lambda)$ is the background value, and $t_{st}$ and $t_{end}$ are the start- and end-time of the flare. $I_{bg}(\lambda)$ was determined as the median of the solar irradiance during 10 minutes before the start of the flare. \\
The integration time is an important parameter for computing the fluence. The end-time of the flare indicated by the GOES flare catalog is the time at which the 0.1--0.8 nm flux comes back to less than half of its value at peak time. This usually underestimates the length of the flare, at least for the hot lines. The starting time of the flare is less difficult to define but we must be sure to include the emission during the impulsive phase. Here again, we considered several definitions of the start- and end-time for our study: 1) we attempted to determine the start- and end-time for each line and each flare in a similar manner to what is done for GOES; 2) the integration is made from 5 minutes before the GOES start-time to 20 minutes after the GOES end-time, and 3) the integration is made between the GOES start- and end-time. \\
Methods 2) and 3) are similar because they assume the same start- and end-time for all lines, but method 2) ensures that no flare emission is missed during the integration. For lines with shorter duration (\textit{e.g.} lines with lower contrast or "impulsive" profile), we expected that the integration over a longer time period would not change the fluence values because the background is removed before the integration. However, this revealed to not be the case, as illustrated in Figure \ref{Fig_TIME}, which shows that integrating over both a too long or too short time with respect to the flare duration leads to an under-estimation of $\gamma$, and therefore an overestimation of the scaling exponent. In the case of a too long integration time, noise will artificially increase the fluence for the smallest flares as explained previously. It is more difficult to explain why this is also the case for a very short interval centered around the peak time, where noise should not play a role; one possible explanation is that using a short and fixed integration time leads to an underestimation of the fluence for the larges flares since, in proportion, less and less of the flare emission is taken into account. Therefore, method 3) was found more reliable.  
 For method 1), the determination of the start time was found to be not robust enough to be used. However, we could determine the end-time of the flare as the time at which the background-removed solar flux reaches 25\% of its peak value or one standard deviation of the time series outside of the flare. The main effect is to increase the flare duration for most wavelengths; this method is however more reliable at wavelengths where the flare contrast is large.
Finally, we have compared the values of the scaling exponent using both method 1) and 3) and the values usually agree within the error bars (see Figures \ref{Fig_Scaling1} and \ref{Fig_Scaling1b} and next section).\\

\begin{figure}    
   \centerline{\includegraphics[width=0.45\textwidth,clip=]{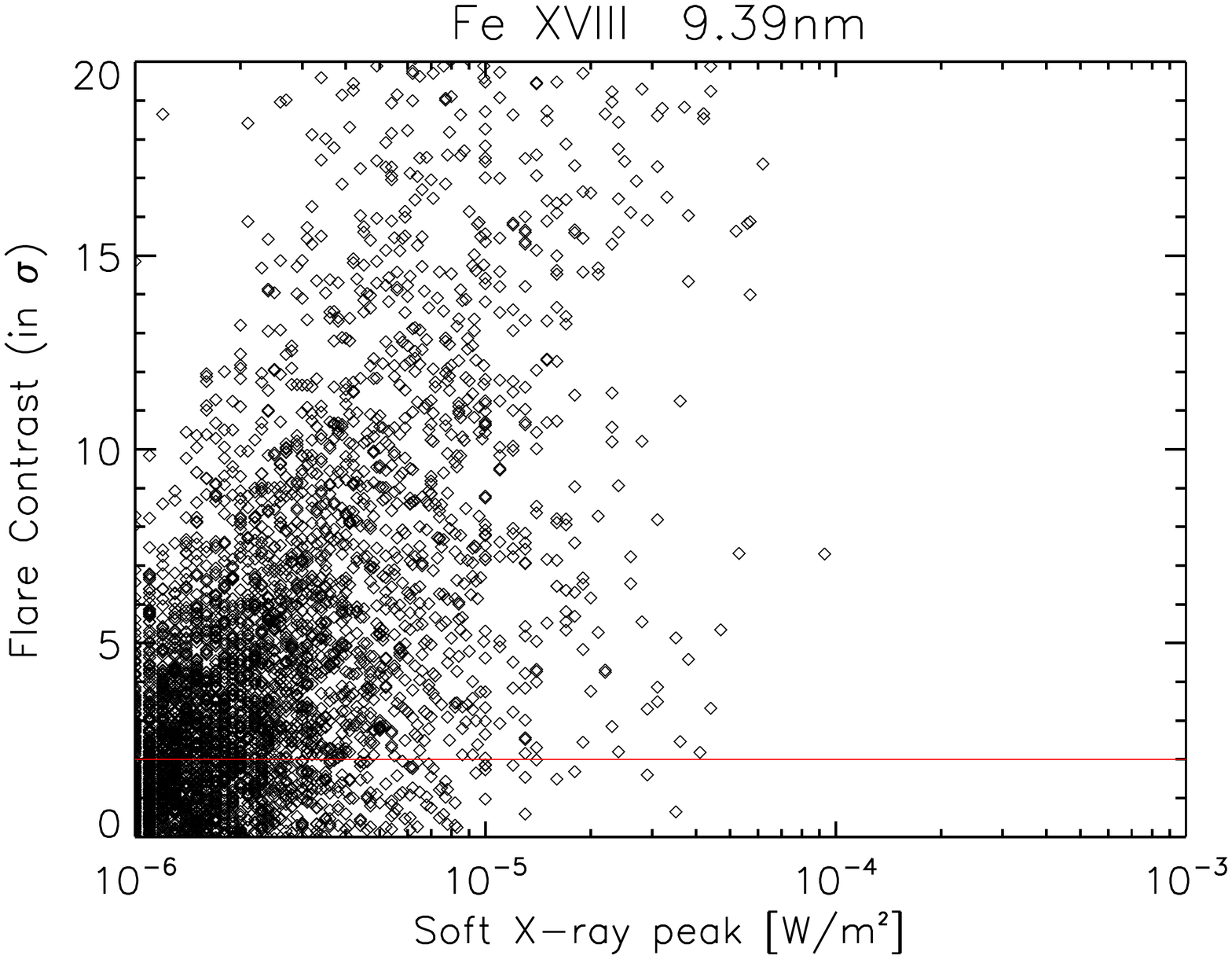} 
   			\includegraphics[width=0.45\textwidth,clip=]{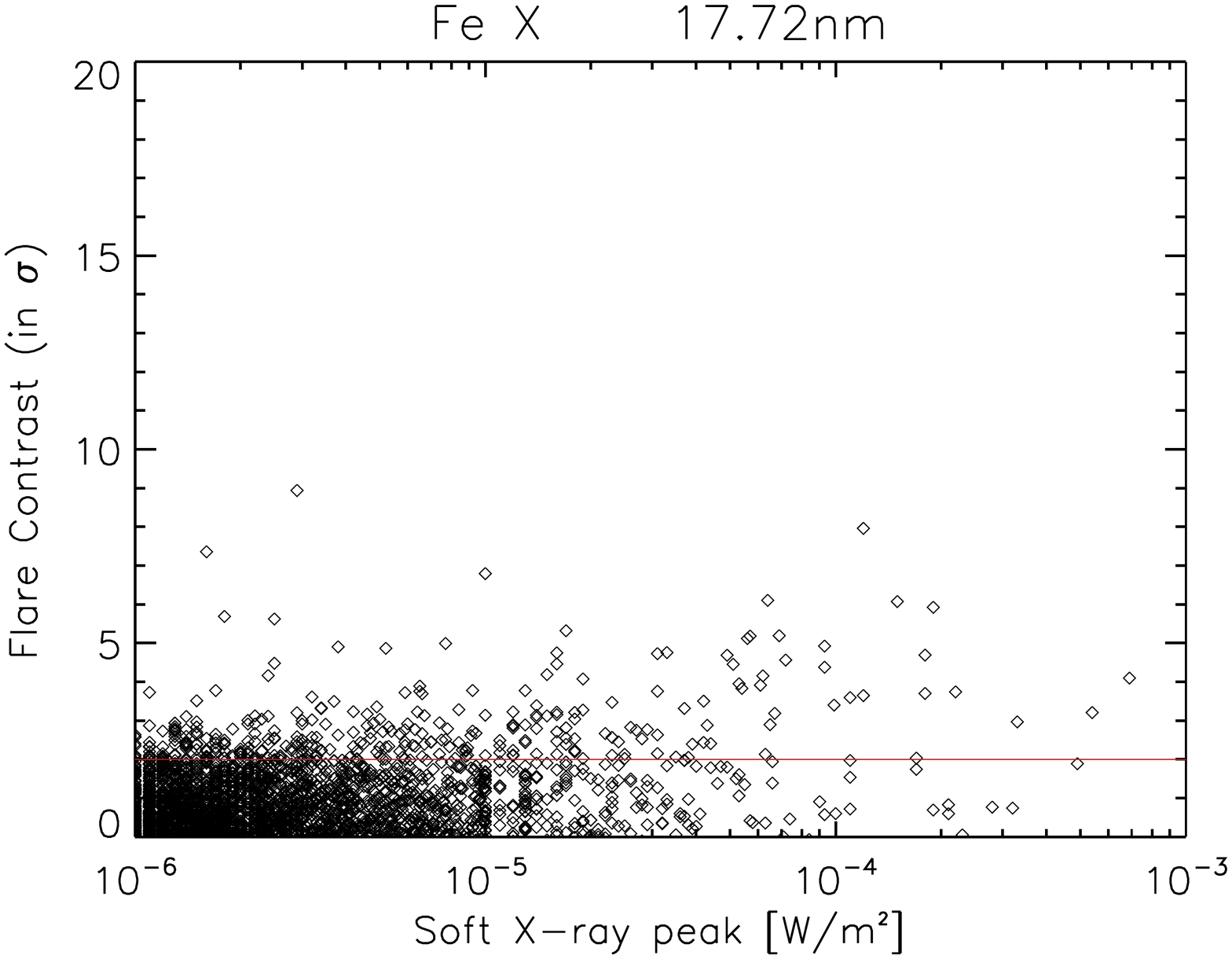} }   
   \centerline{\includegraphics[width=0.45\textwidth,clip=]{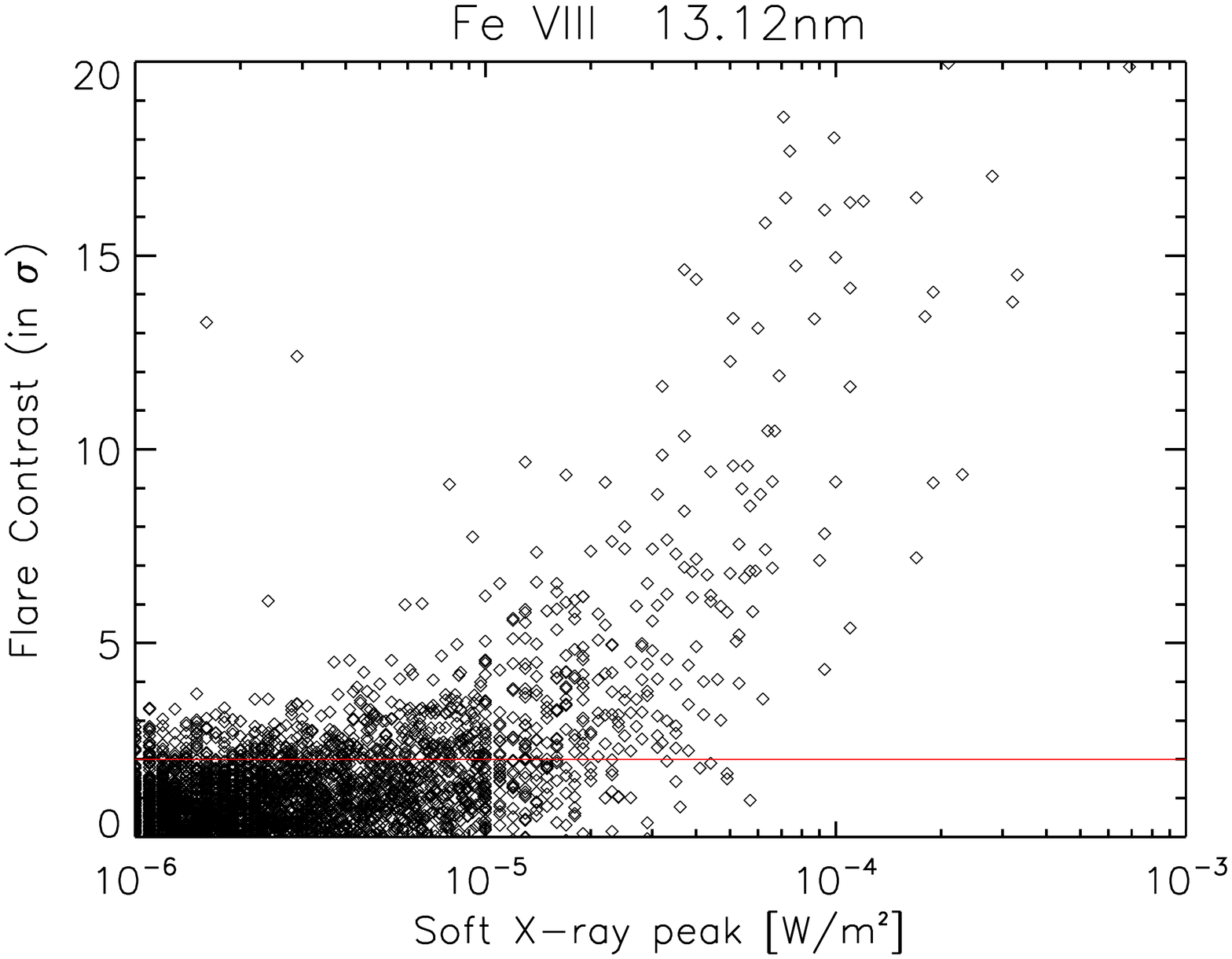} 
   			\includegraphics[width=0.45\textwidth,clip=]{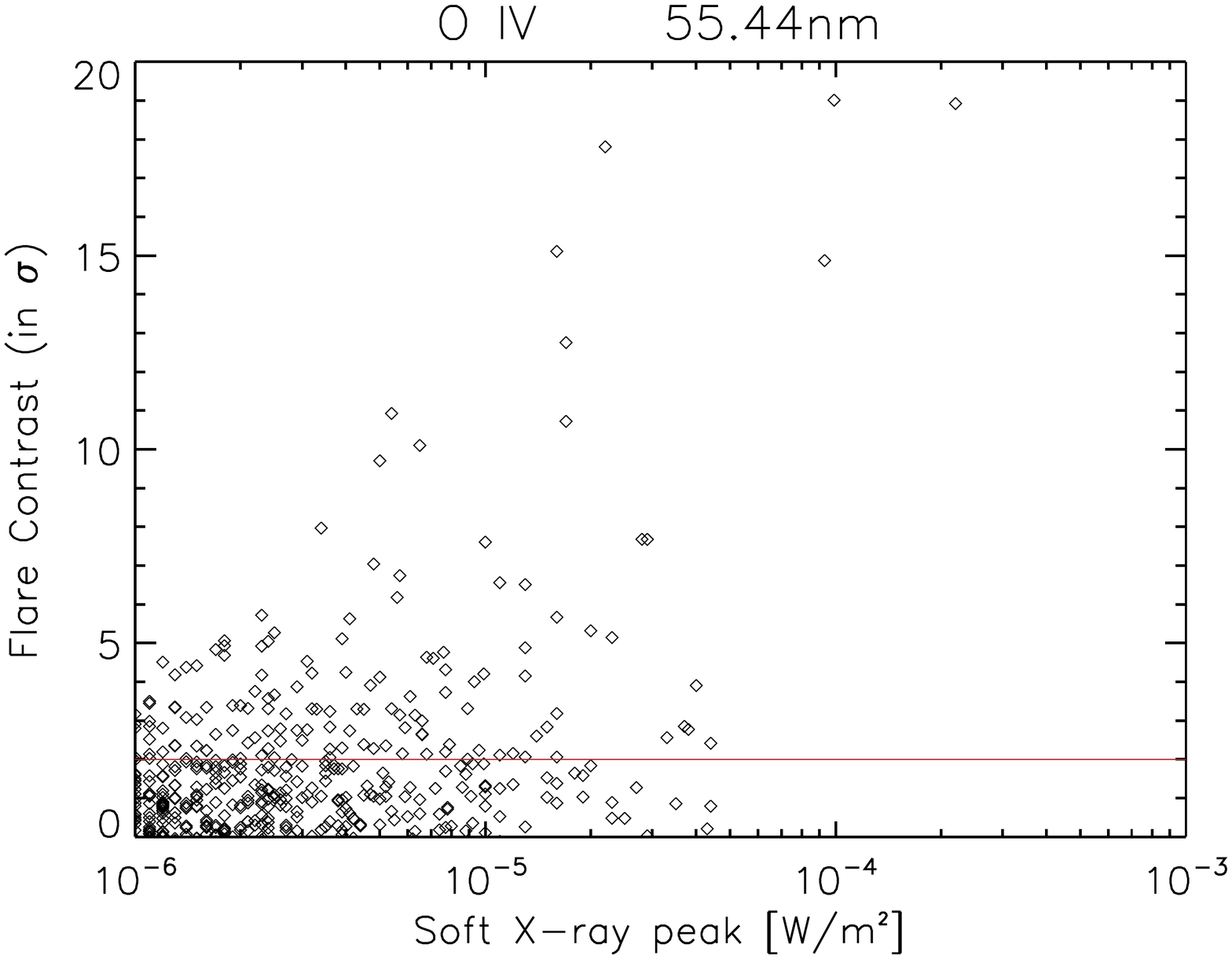} }   
               \caption{ Signal to noise ratio of the flare fluence (denoted flare contrast in the figure) for four lines. In each panel, a red line is drawn at the 2-sigma level and only flare above this level were considered.}
   \label{Fig_SNR}
   \end{figure}

We checked that not other flares occur during the integration time and used only line fluences that have a signal to noise ratio (SNR) larger than two sigma. The signal to noise ratio  $S(E)$ of the fluence $E$ for each flare and line is defined as
$$ S(E)=\frac{E}{\int_{t_{st}}^{t_{end}} \sigma_{preflare} dt} ,$$
where $\sigma_{preflare}$ is the standard deviation of the time series measured before the flare. We required that $S(E) > 2$ to consider the flare fluence in the statistics. Figure \ref{Fig_SNR} shows the SNR for different lines. One can see that basically only flares above the M-class can be used. While the fluence of the Fe XVIII line at 9.39 nm has generally a good SNR, the Fe X line at 17.22 nm cannot be used at all. The set of lines that we used and that is shown in Table \ref{T-Lines} is based on the criteria ($S(E) > 2$) and on the requirement that the fluxes computed with fixed and non-fixed spectral limits are in good agreement. We are interested in chromopsheric (low temperature) UV lines but they have usually a relatively small contrast and basically show up clearly only for M-flares and above. In order to compare meaningfully different lines, all the scaling are done on flares above the M1 level in the following, although C-class flares are also shown in the plots.

%
%

\begin{figure}    
   \centerline{
   \includegraphics[width=0.45\textwidth,trim = 0.65cm 2.3cm 1.1cm 2.1cm,clip=]{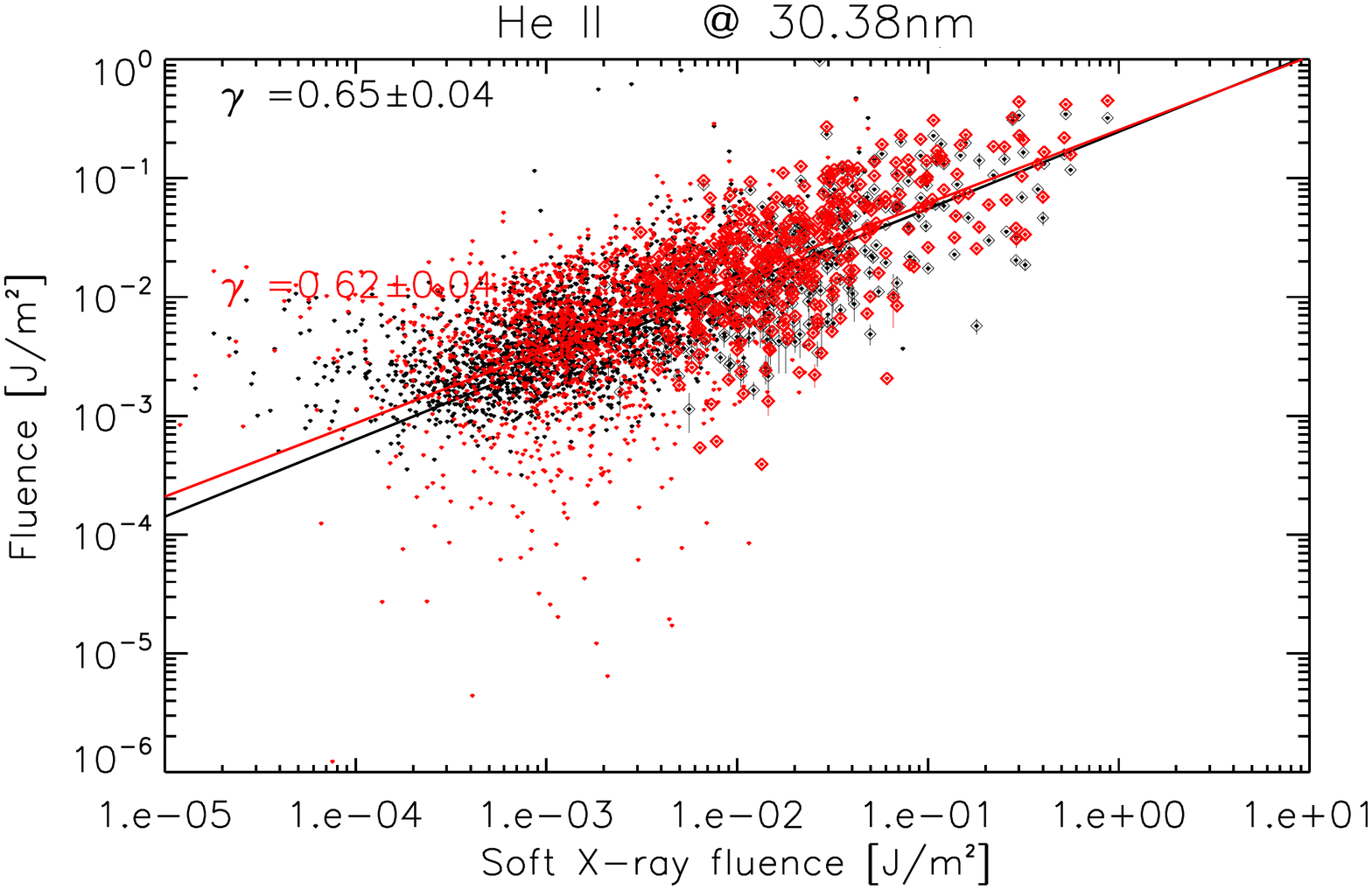}
  \includegraphics[width=0.45\textwidth,trim = 0.65cm 2.3cm 1.1cm 2.1cm,clip=]{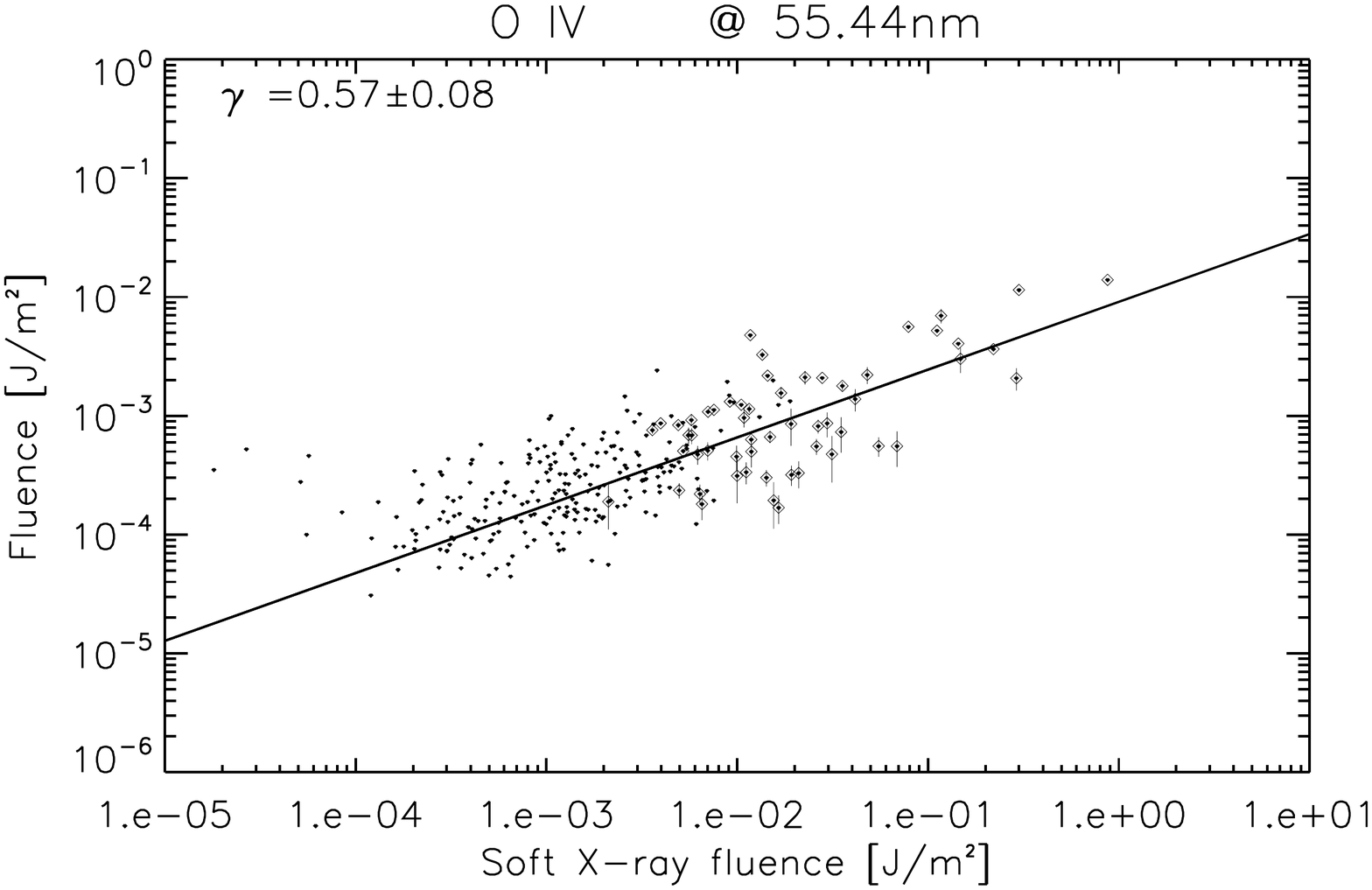} 
   }

   \centerline{
   \includegraphics[width=0.45\textwidth,trim = 0.65cm 2.3cm 1.1cm 2.1cm,clip=]{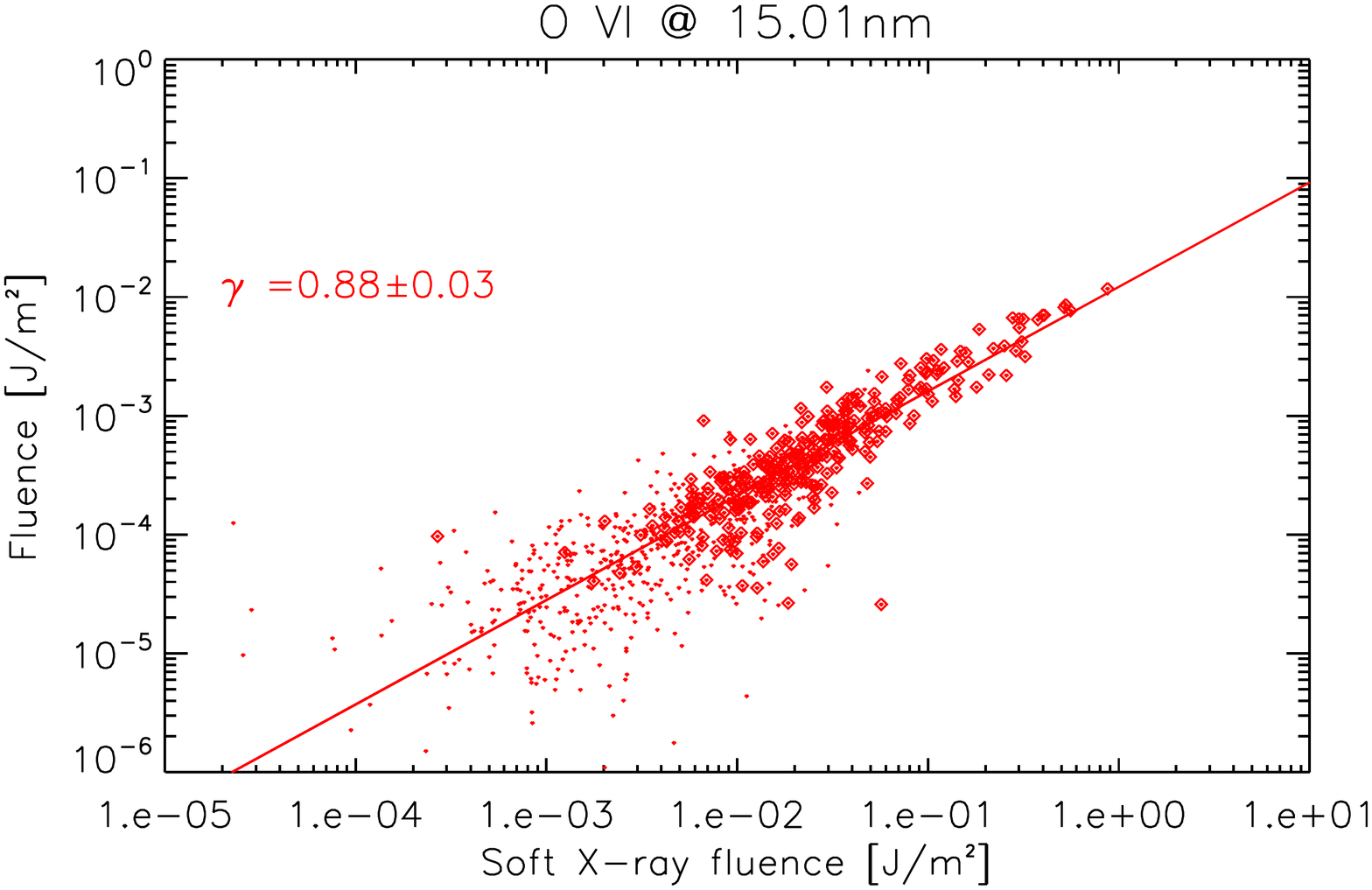}
  \includegraphics[width=0.45\textwidth,trim = 0.65cm 2.3cm 1.1cm 2.1cm,clip=]{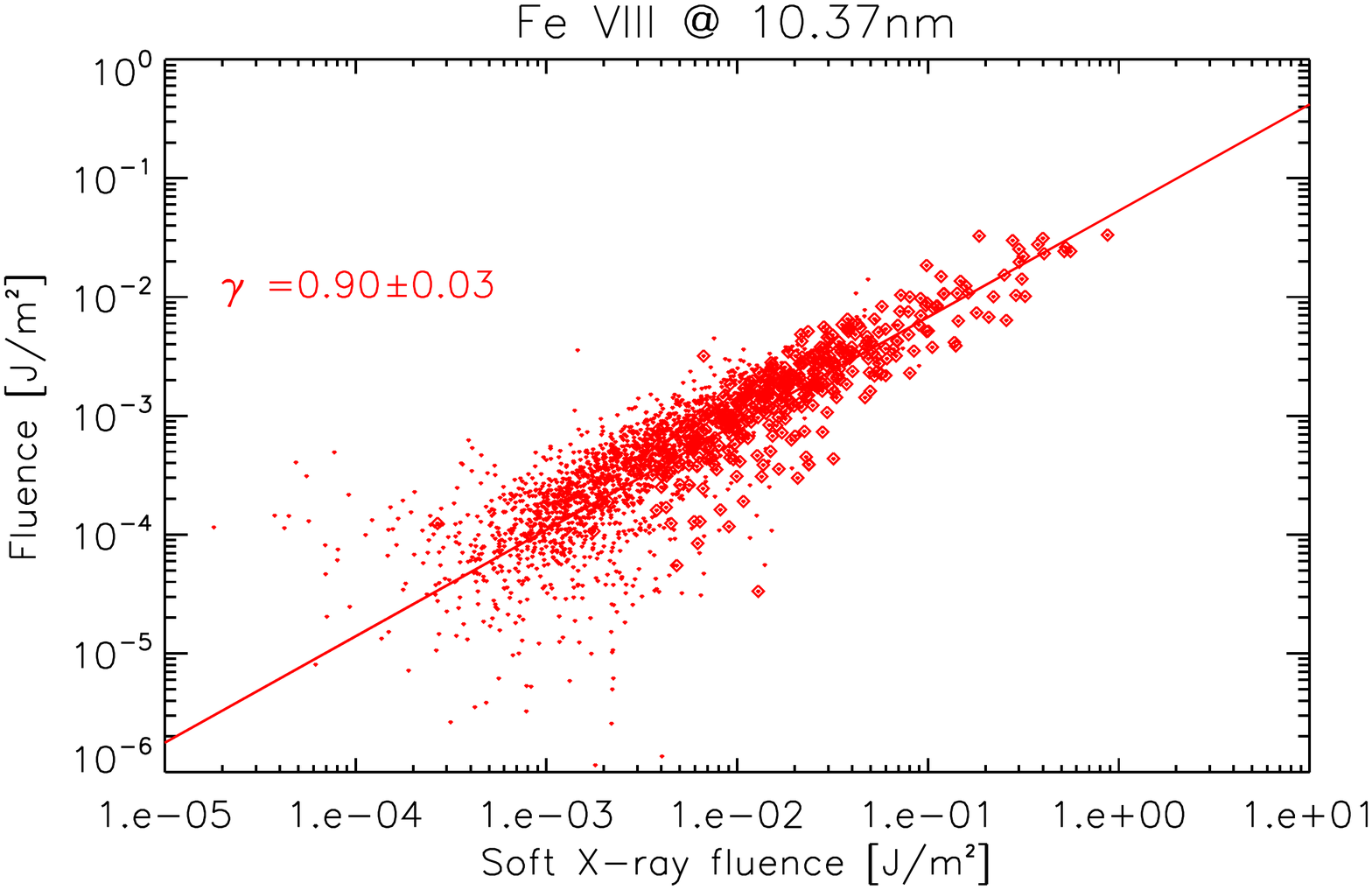}
}
   \centerline{
  \includegraphics[width=0.45\textwidth,trim = 0.65cm 2.3cm 1.1cm 2.1cm,clip=]{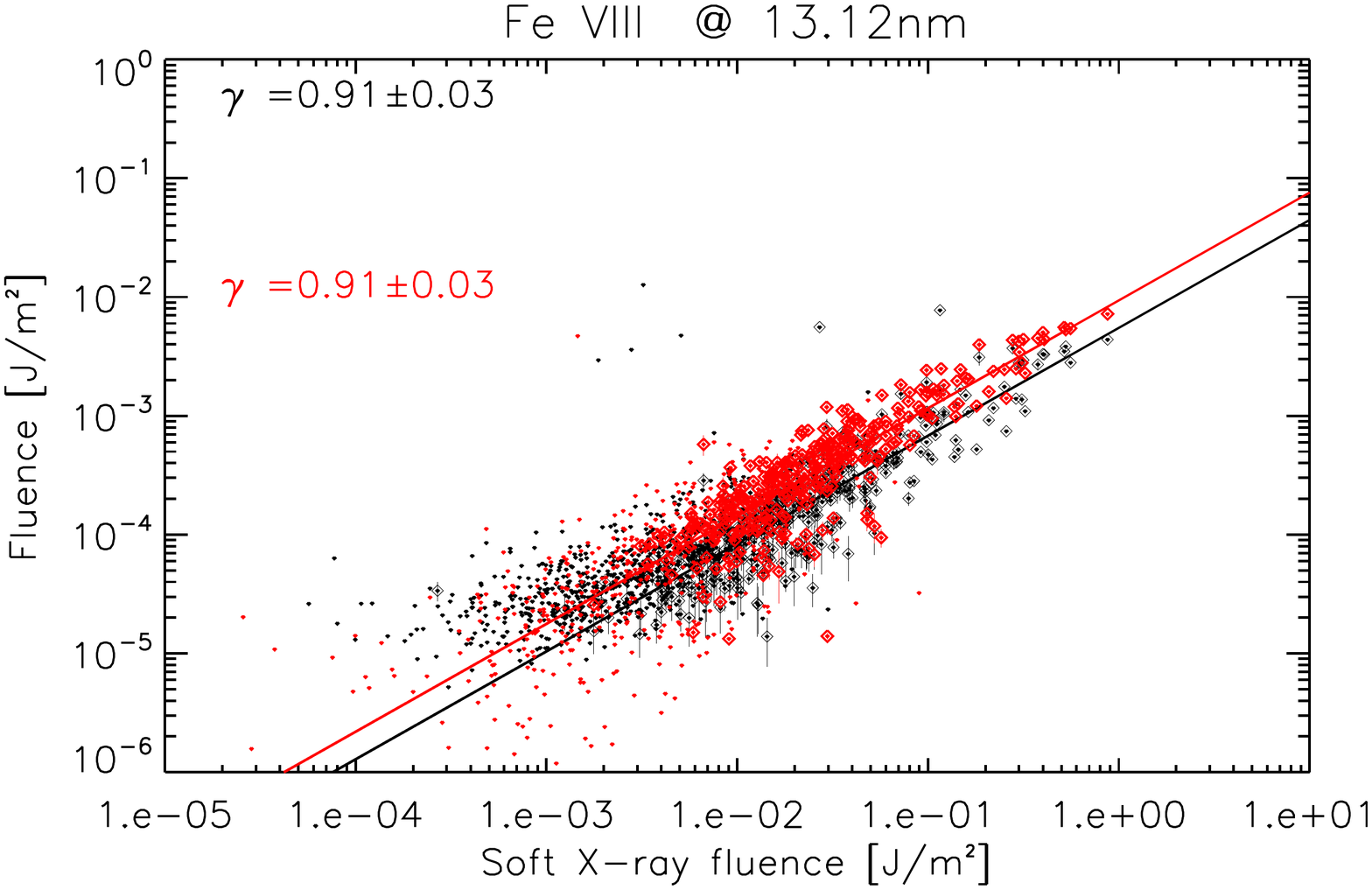} 
   \includegraphics[width=0.45\textwidth,trim = 0.65cm 2.3cm 1.1cm 2.1cm,clip=]{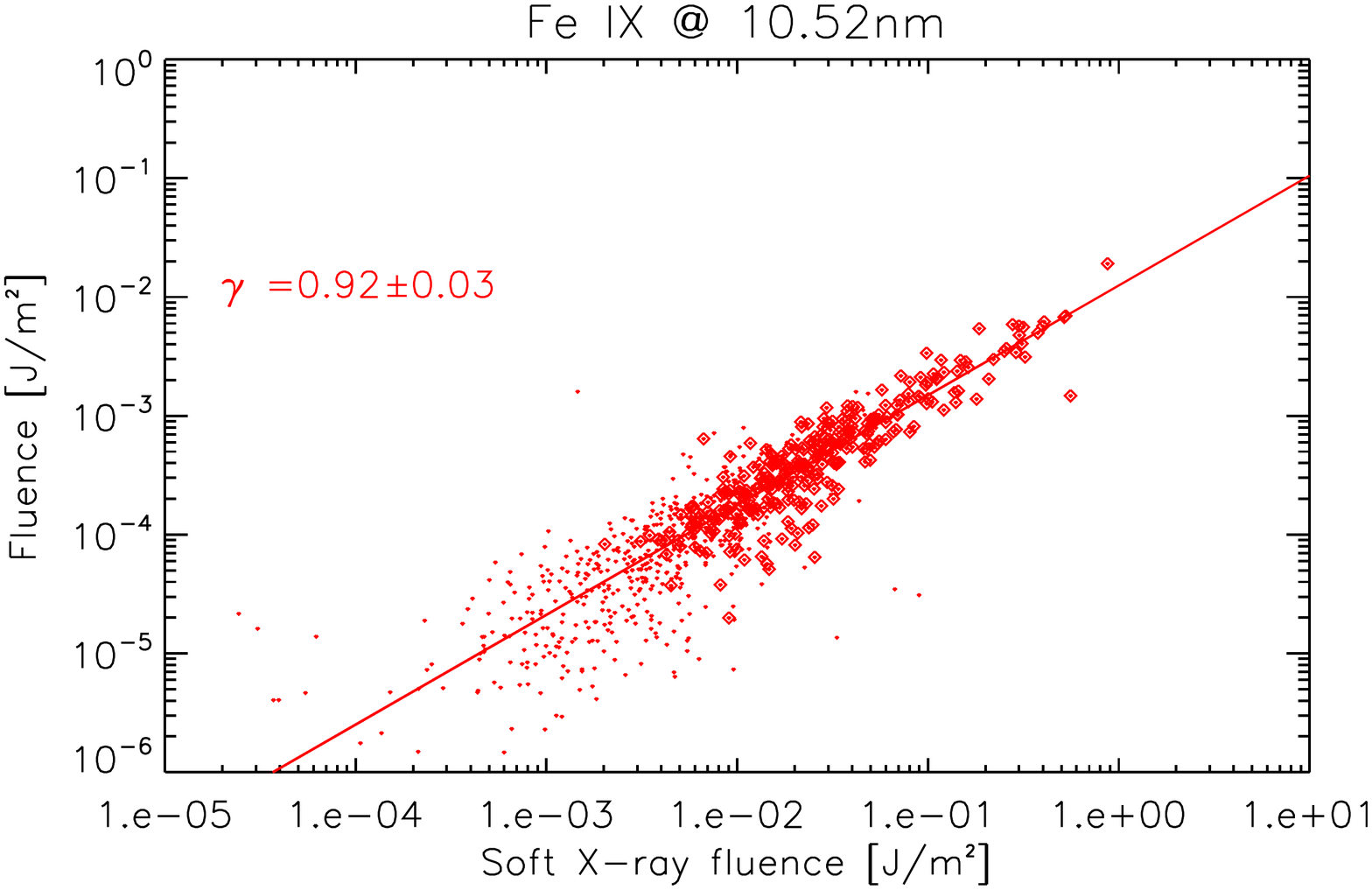}
   }
 
   \centerline{
  \includegraphics[width=0.45\textwidth,trim = 0.65cm 2.3cm 1.1cm 2.1cm,clip=]{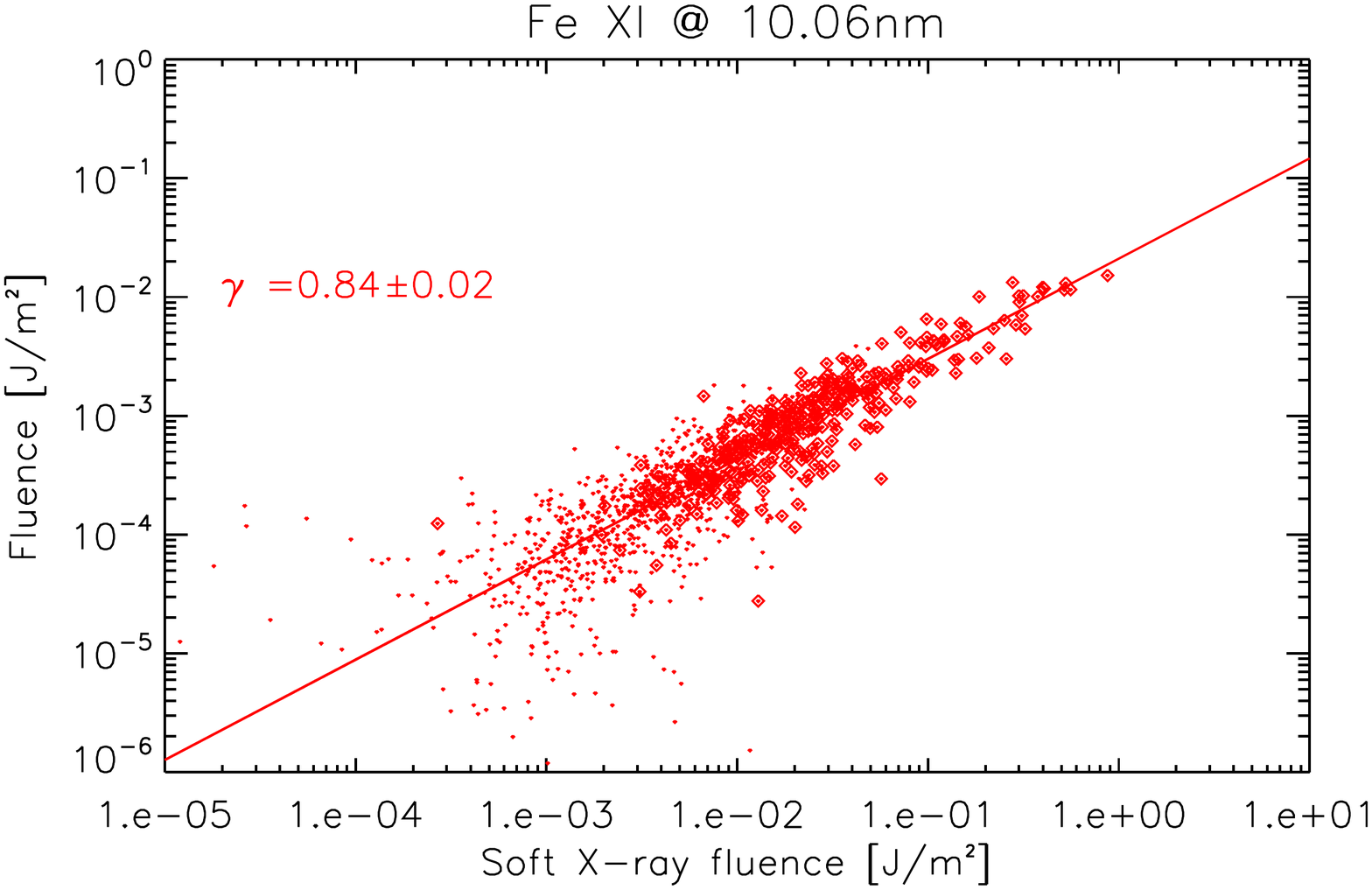}
   \includegraphics[width=0.45\textwidth,trim = 0.65cm 2.3cm 1.1cm 2.1cm,clip=]{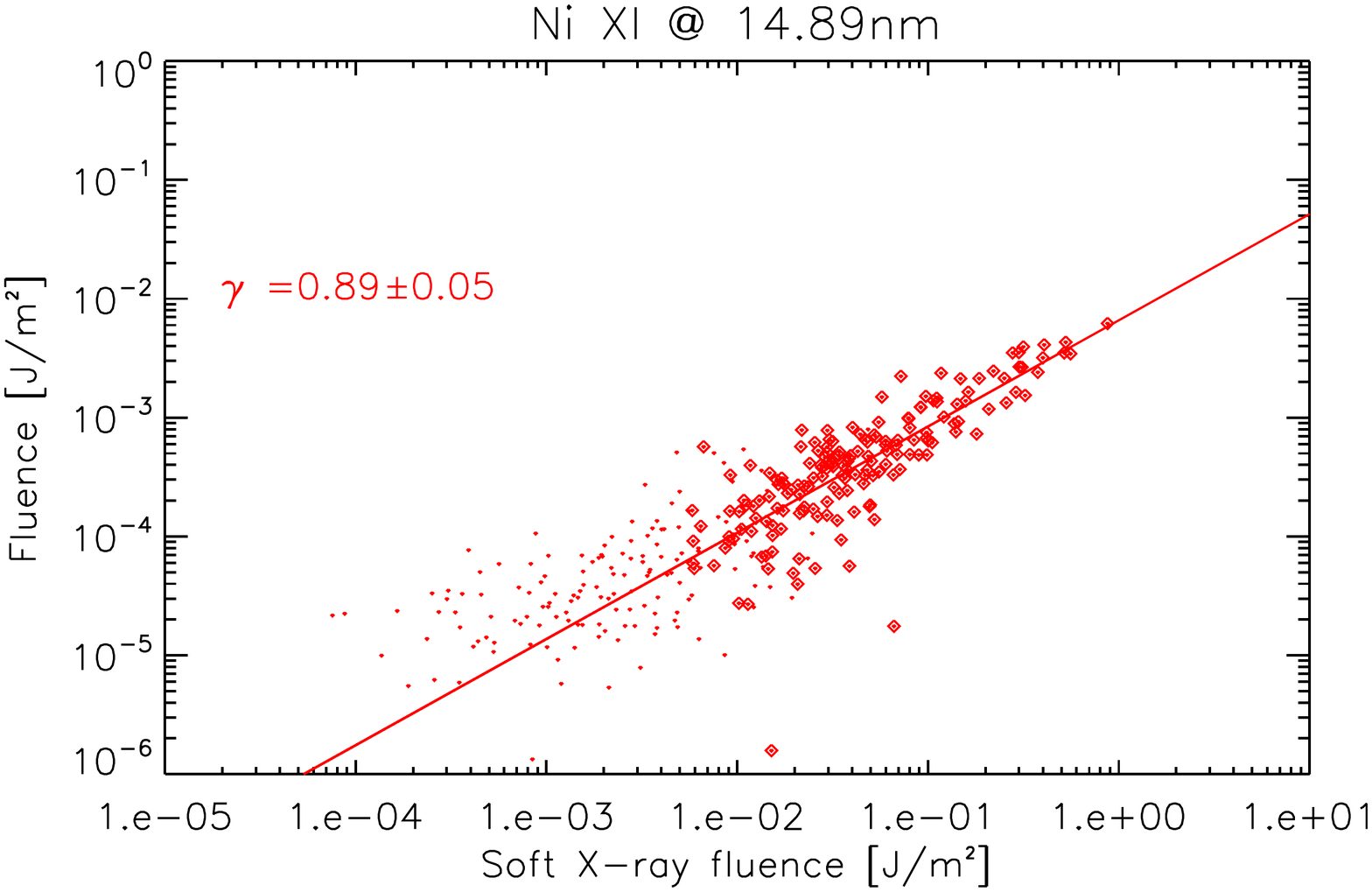}
}
   \centerline{
   \includegraphics[width=0.45\textwidth,trim = 0.65cm 2.3cm 1.1cm 2.1cm,clip=]{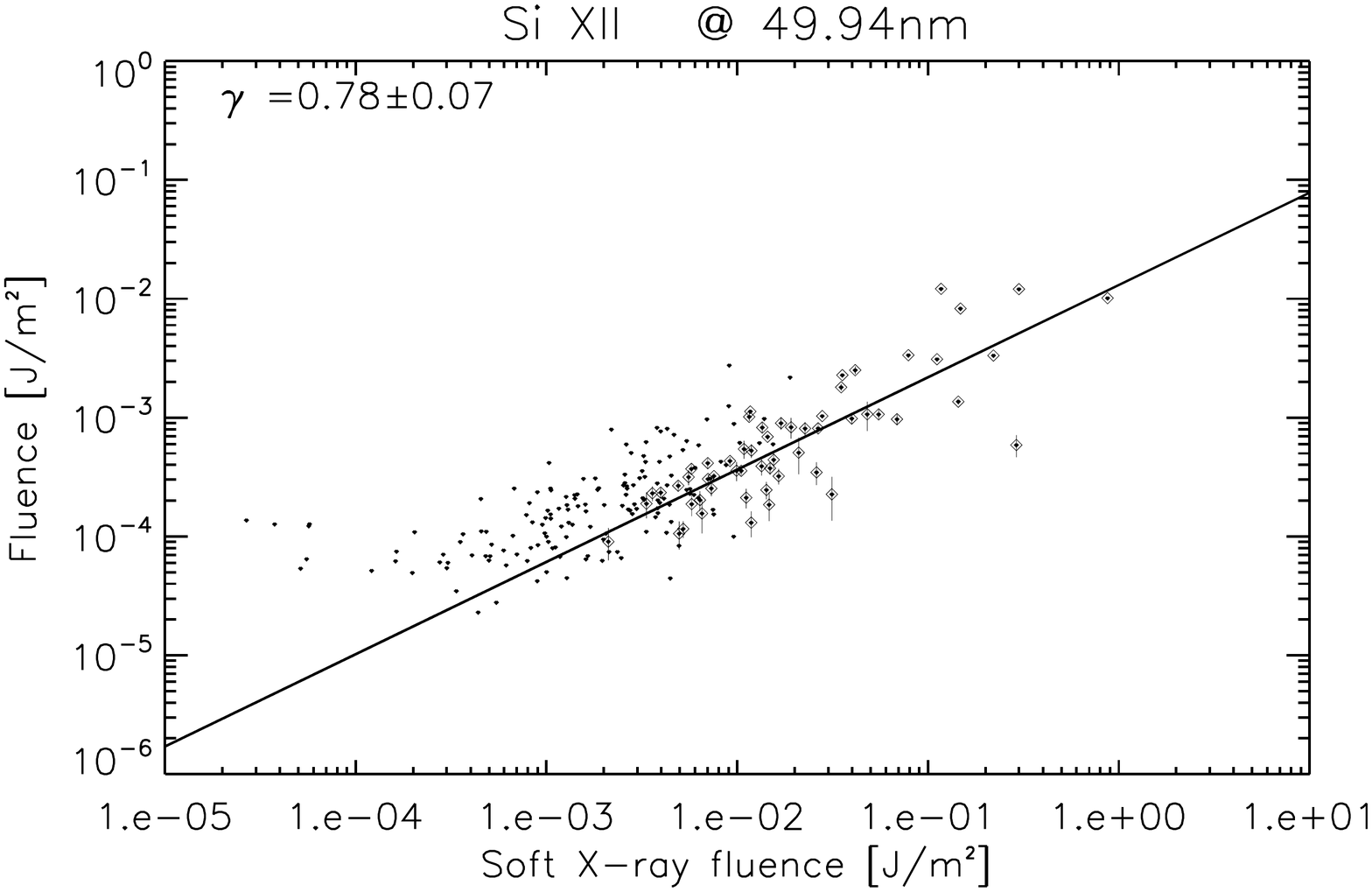}
   \includegraphics[width=0.45\textwidth,trim = 0.65cm 2.3cm 1.1cm 2.1cm,clip=]{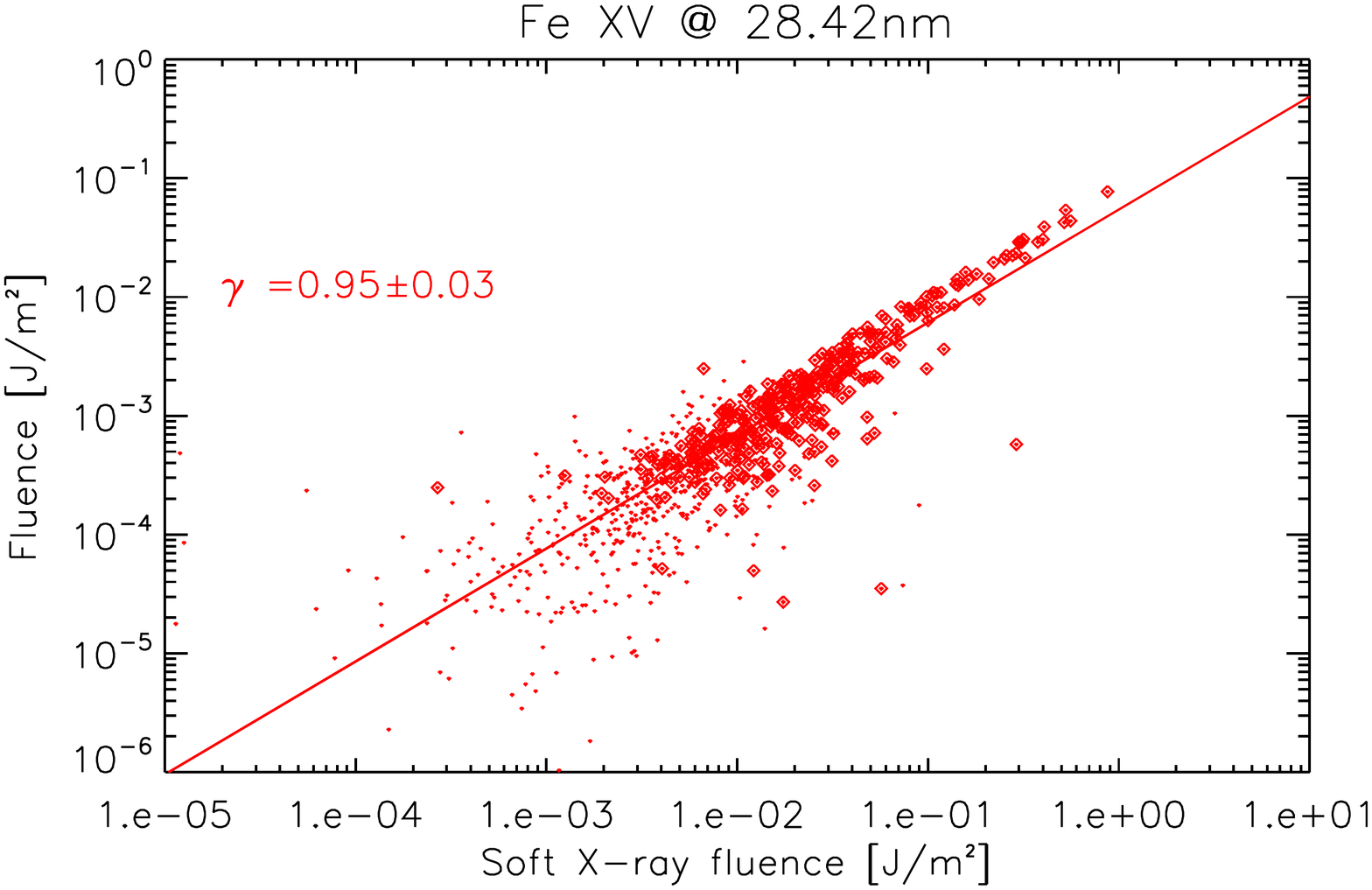}
}

           \caption{ \textbf{(continued on Figure \ref{Fig_Scaling1b})} Scaling of the flare fluence for different lines versus the SXR GOES flare fluence. Black color corresponds to fluences computed with the GOES start- and end-time and with the EVE lines product ("fixed spectral integration"). Red corresponds to fluences computed by automatically determining the line spectral width and the end of the flare. The fits are performed on the flares above the M1 level, represented with larger diamonds. For clarity, error bars are drawn only for flares above the M1 level. }
   \label{Fig_Scaling1}
\end{figure}

\begin{figure}    

   \centerline{
  \includegraphics[width=0.45\textwidth,trim = 0.65cm 2.3cm 1.1cm 2.1cm,clip=]{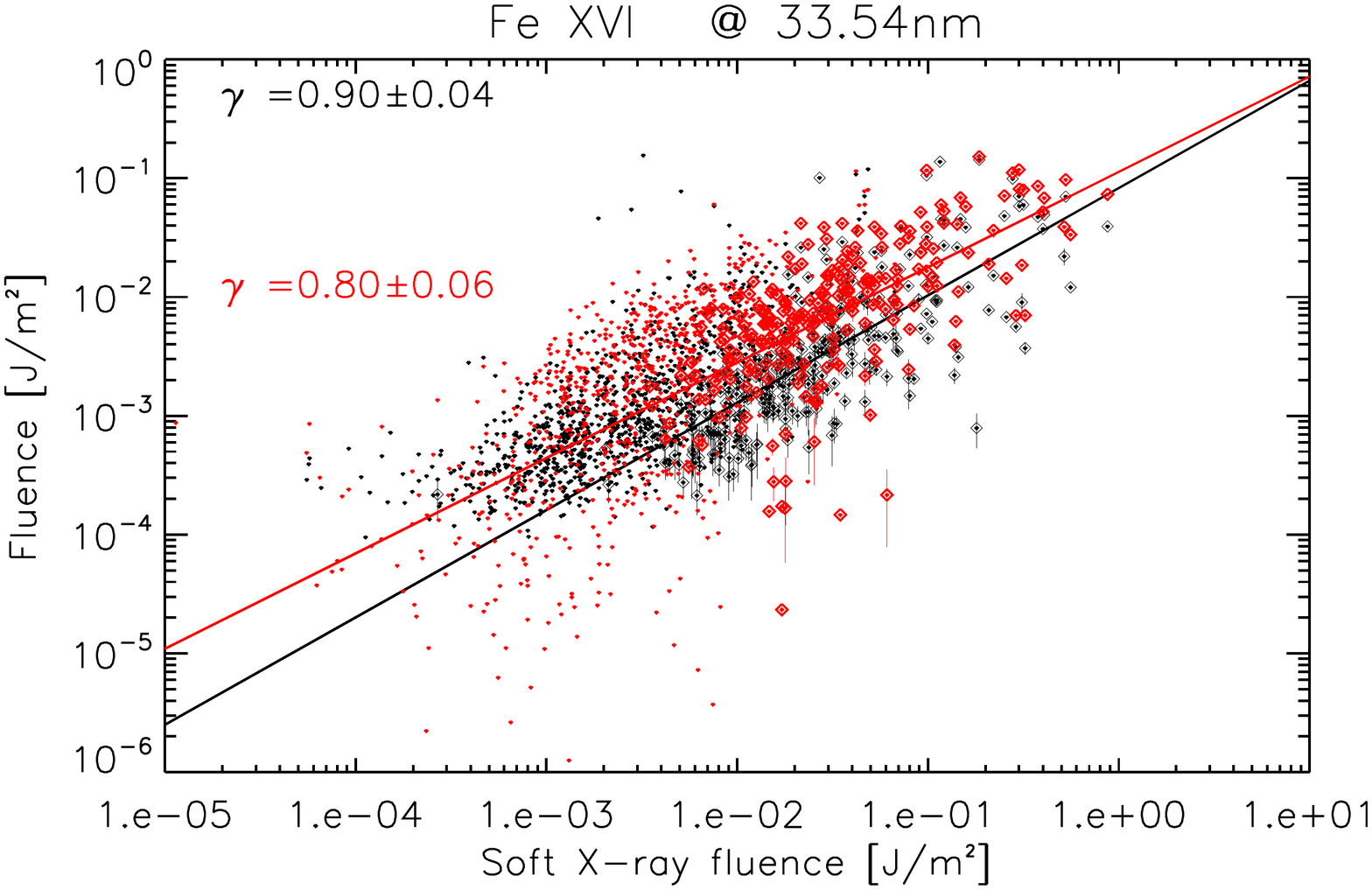}
  \includegraphics[width=0.45\textwidth,trim = 0.65cm 2.3cm 1.1cm 2.1cm,clip=]{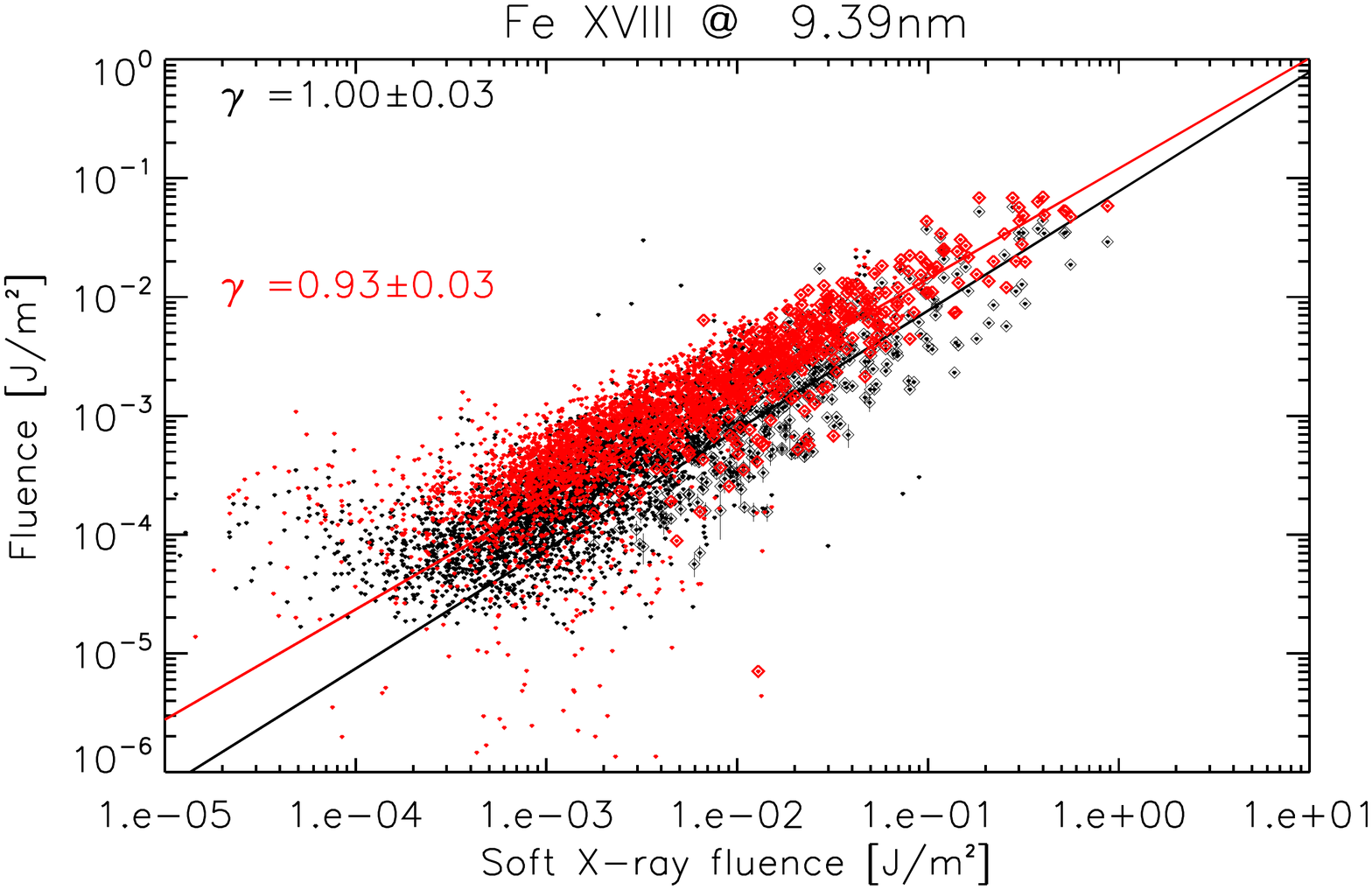}  
  }

   \centerline{
   \includegraphics[width=0.45\textwidth,trim = 0.65cm 2.3cm 1.1cm 2.1cm,clip=]{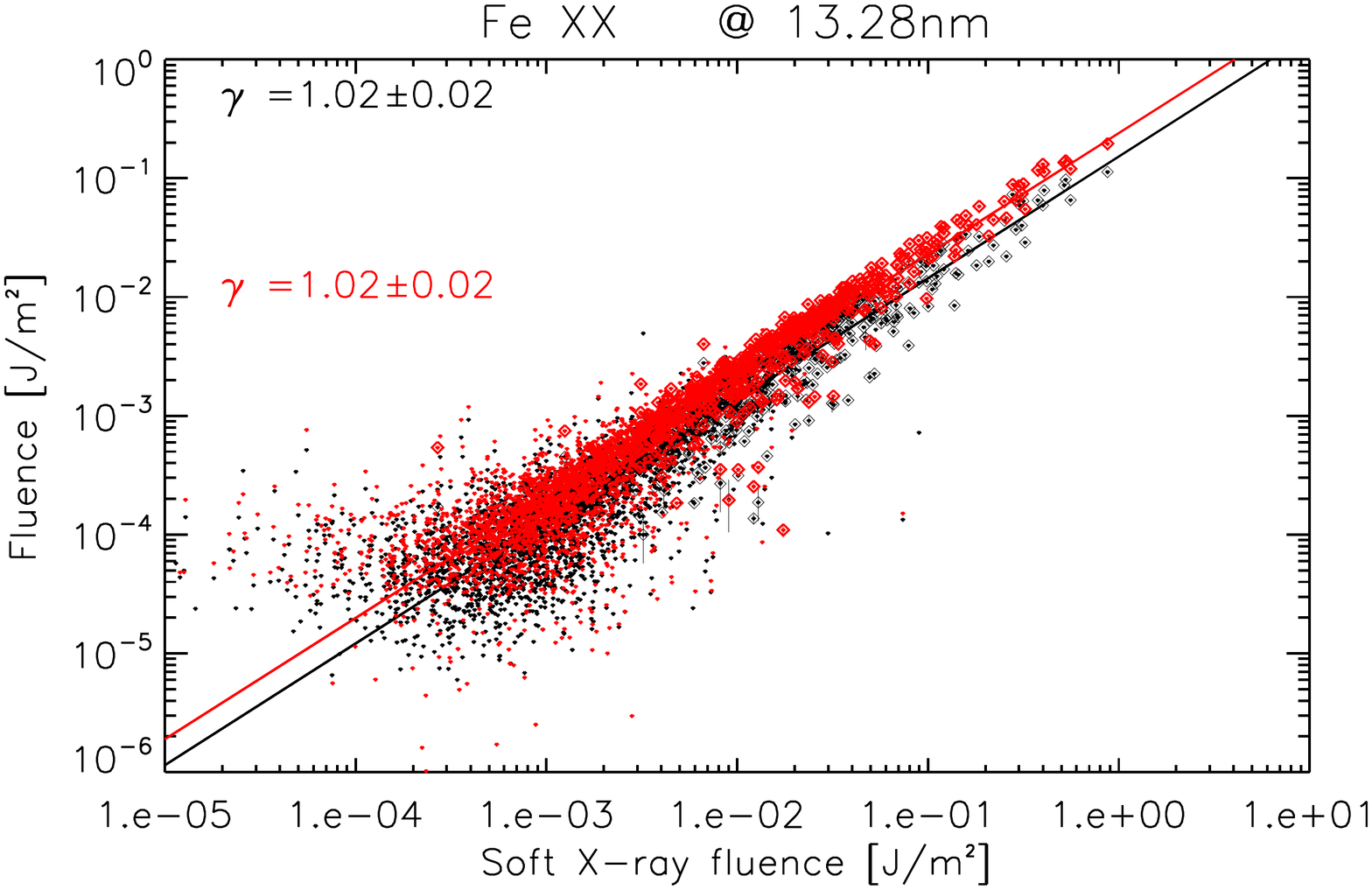}
  \includegraphics[width=0.45\textwidth,trim = 0.65cm 2.3cm 1.1cm 2.1cm,clip=]{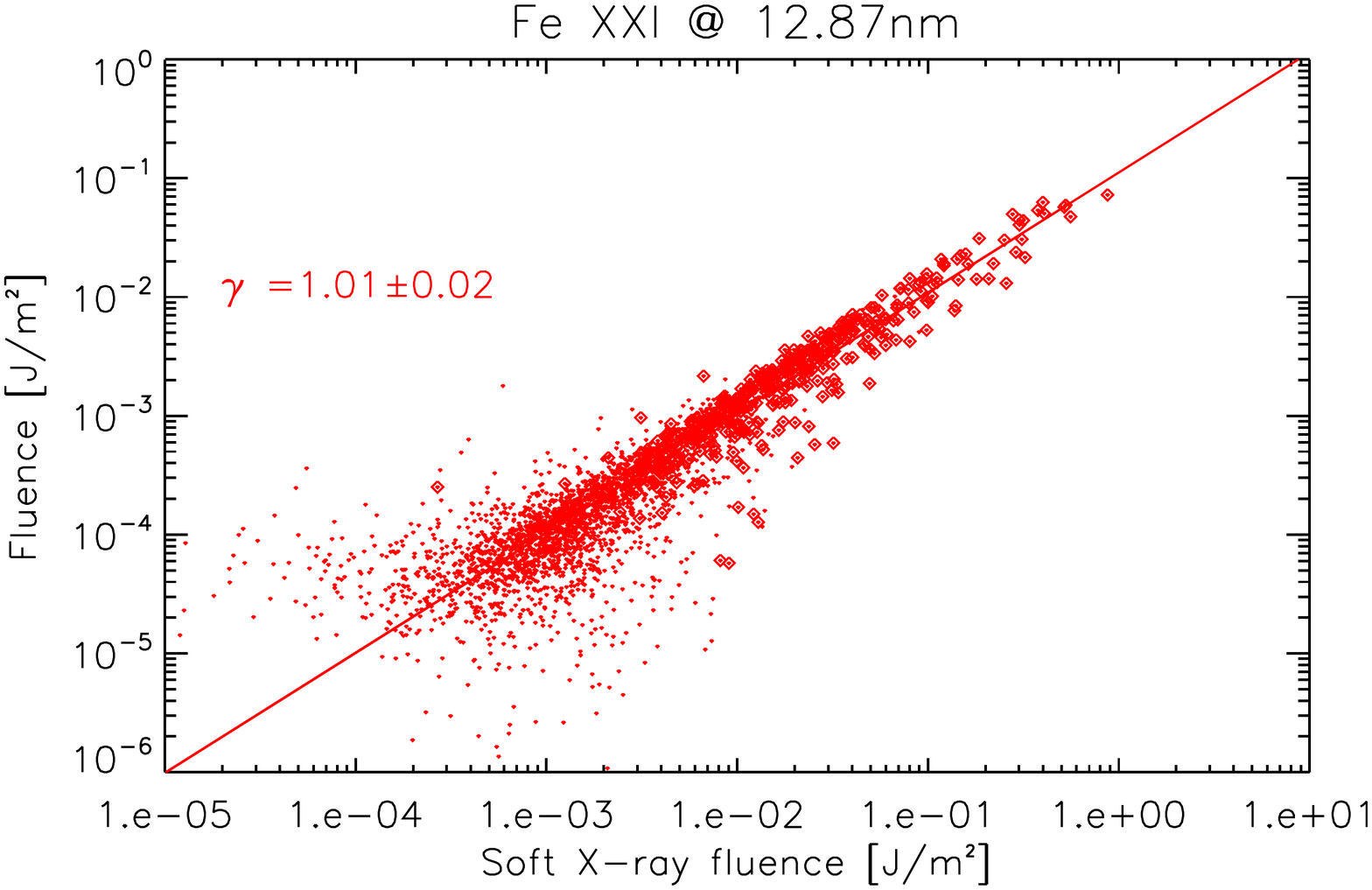}
}
   \centerline{  \includegraphics[width=0.45\textwidth,trim = 0.65cm 2.3cm 1.1cm 2.1cm,clip=]{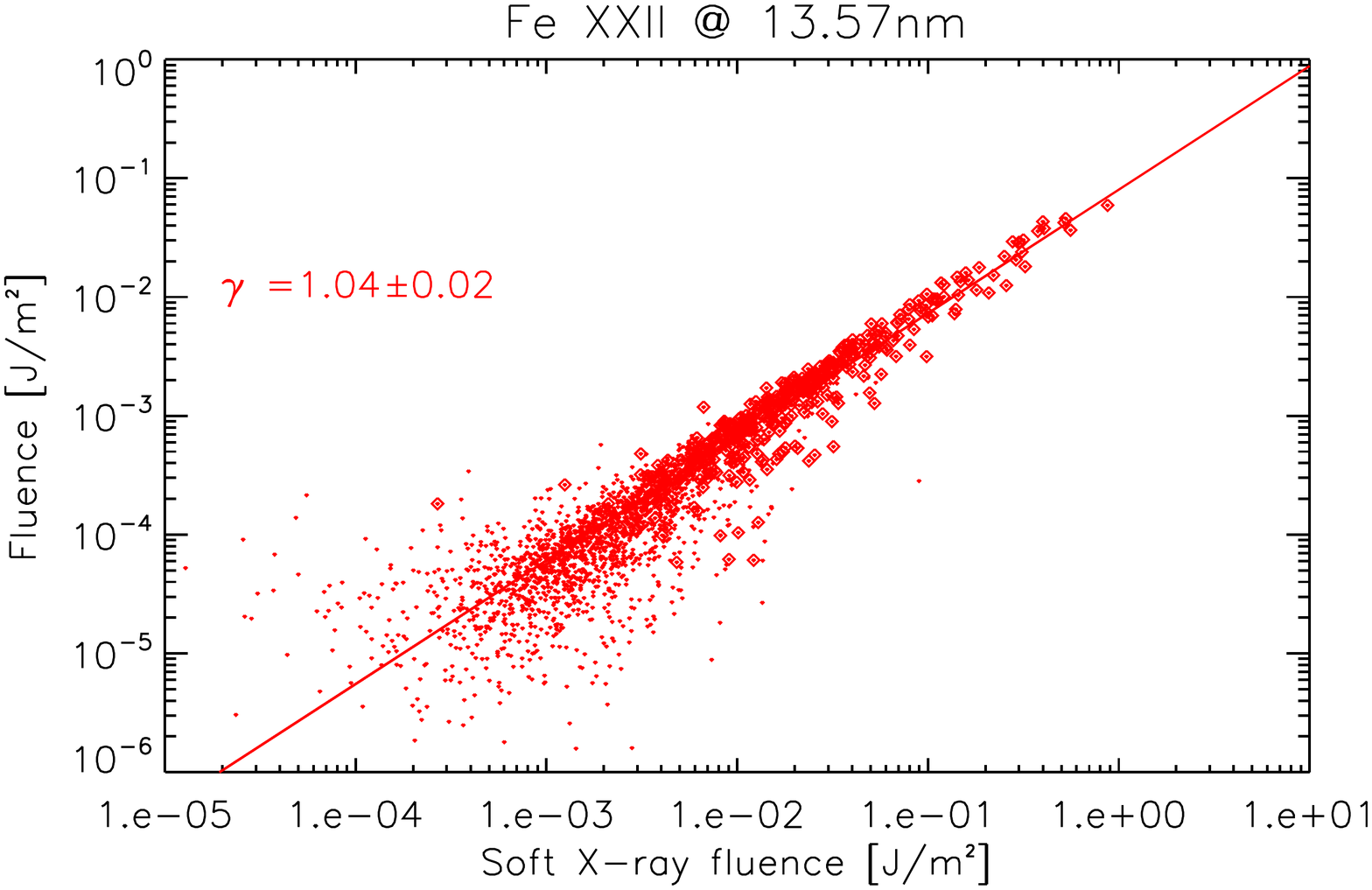} 
   }

           \caption{ Following of Figure \ref{Fig_Scaling1}}
   \label{Fig_Scaling1b}
   
\end{figure}             


\begin{figure}    
   \centerline{\includegraphics[width=0.8\textwidth,clip=]{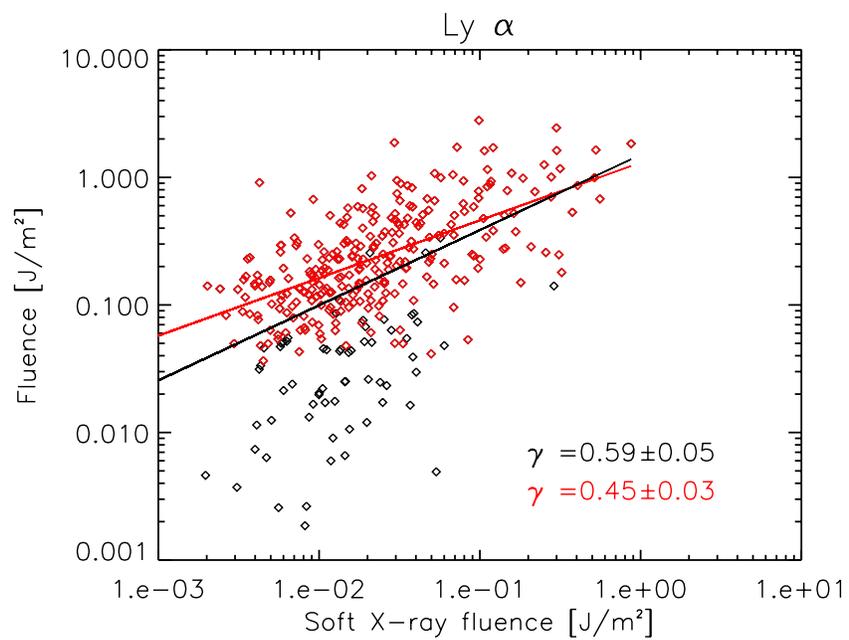} 
                 }
              \caption{Scaling of the Ly-$\alpha$ fluence with the GOES SXR fluence. The straight lines show the fit on respectively all flares above the M1 level (in black) and flares with SNR $>$ 2$\sigma$ (in red). }
   \label{Fig_lya}
   \end{figure}

\section{Results and Discussion} 
      \label{S-results}

\begin{figure}    
   \centerline{\includegraphics[width=0.8\textwidth,clip=]{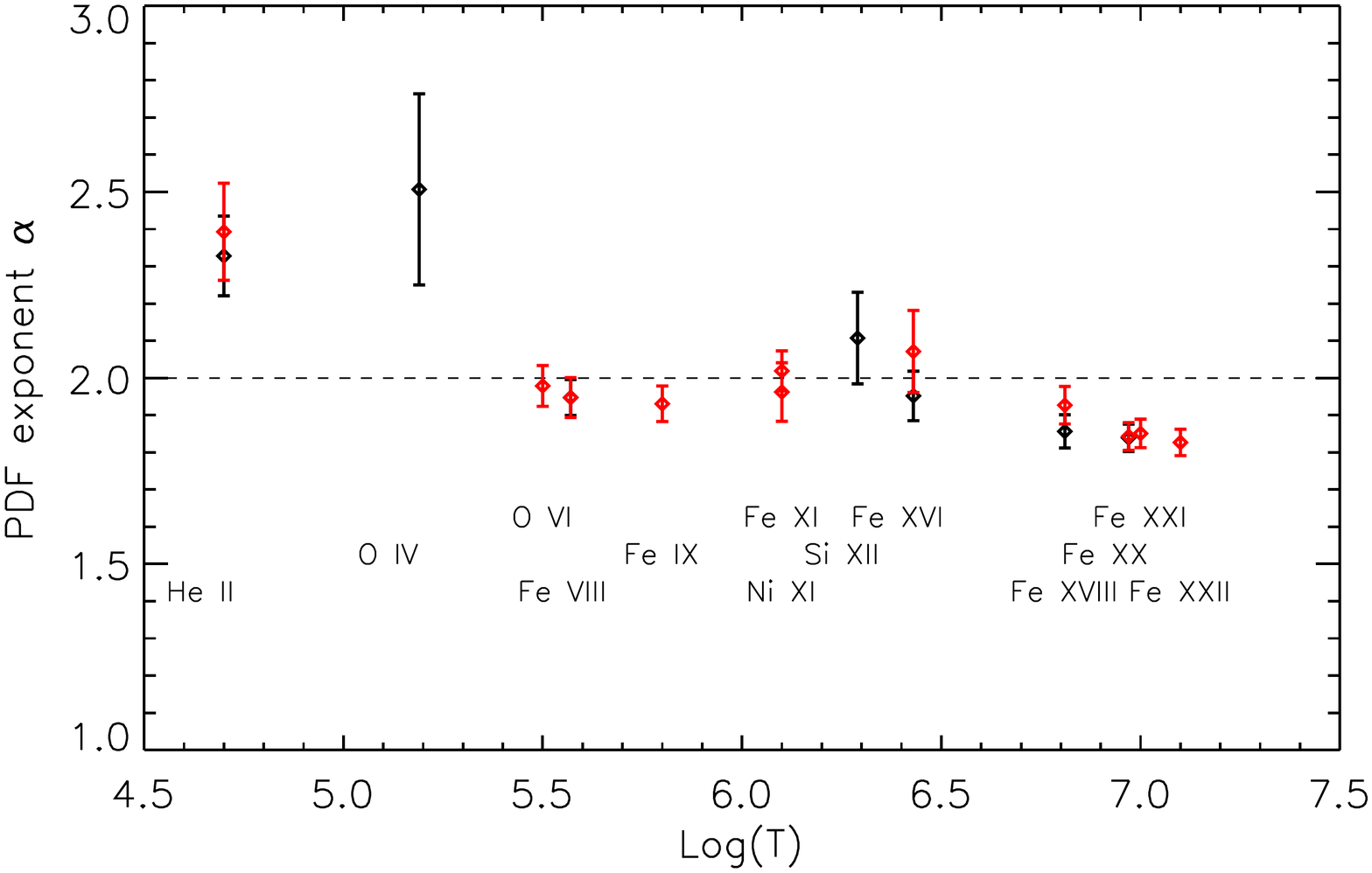} 
                 }
              \caption{Scaling exponent of the fluence PDF versus  temperature.}
   \label{Fig_Expo}
   \end{figure}

Figures \ref{Fig_Scaling1} and \ref{Fig_Scaling1b} show the scaling of the flare fluence for 16 spectral lines, and the determination of the $\gamma$ exponent, such that $E(\lambda)\simeq E_{SXR}^{\gamma}$. The black diamonds represents flare fluence computed using the EVE spectral lines product with fixed spectral integration limits and all flares with $S(E) > 2$ are plotted. Red diamonds are the fluences that we computed by determining for each spectrum the blue and red limits of the integration for a spectral line. Only flares with $S(E) > 2$ and that have a SXR class larger than M1.0 are plotted. In both cases, the fit is performed on flare above the M1 level only, to avoid taking into account flares with too small contrast and that does not appear to belong to the distribution. \\
Some lines have larger scatter than others; this is the case for the Fe XV line at 28.42 nm and the Fe XVI line at 33.54 nm. Some lines also have a smaller number of "valid" flares : O IV at 55.44 nm, Si XII at 49.94 nm, both observed by MEGS-B. Let us note that because of several wavelengths are missing in the MEGS B spectrum product, we could not validate the fluences computed from the EVE line product for these lines by reproducing them with the spectrum product. The scattering and small number of points for the regression are reflected in the 1-$\sigma$ uncertainty of $\gamma$, which is then used to compute the uncertainty on $\alpha$ represented in Figure \ref{Fig_Expo}. From the top left panel of Figure \ref{Fig_Scaling1} to the bottom panel of of Figure \ref{Fig_Scaling1b},  there is a small tendency for the scaling to steepen; in other world,  $\gamma$ is slightly increasing or the scaling exponent of the PDF $\alpha$ is slightly decreasing with the temperature of formation of the lines.  \\
Figure \ref{Fig_lya} shows similar result for the H I Lyman-$\alpha$ line. We used the same methods than for the other lines and applied them on the 1-minute observations of the \textit{Extreme Ultraviolet Sensor} (EUVS) onboard the GOES 15 satellites \citep{Viereck:2007fk}. The channel E of the GOES/EUVS instrument is a broad passband centered around Ly-$\alpha$. Because of strong degradation, the absolute value of this channel is scaled to the value of a 1 nm bin around Ly-$\alpha$ as observed by SORCE/SOLSTICE. The scaling is made on the basis of a quiet Sun reference spectrum, which may affect the flare fluence. This channel also suffers from geocoronal hydrogen absorption a few hours per day. These effects may partly explain the scatter observed on Figure \ref{Fig_lya}, but the scatter could also reflect the complexity of the formation of the Ly-$\alpha$ line, which is the most intense line of the solar spectrum. More detailed analysis of Ly-$\alpha$ flare profiles are worthwhile, but Figure \ref{Fig_lya} already shows a reasonable scaling of its fluence with the GOES SXR fluence. The values of $\gamma$ indicated on the figure lead to a value of $\alpha$ between 2.3 and 2.9.\\
 
Figure \ref{Fig_Expo} shows the evolution of the scaling exponent with the formation temperature of the lines (Lyman-$\alpha$ is not represented in this figure). Error bars represent the 1-$\sigma$ uncertainties on $\alpha$ values and are computed using eq.\ref{Eq-expo-Unc} and the 1-$\sigma$ uncertainties of $\gamma$ values coming from the fitting procedure. The O IV line at 55.44 nm (Log(T)=5.5) and Si XII (log(T)=6.3) have larger error bars, in agreement with the small number of flares and the large scatter observed in Figures \ref{Fig_Scaling1} and Figure \ref{Fig_Scaling1b}. More precise results are obtained without surprised for the hottest and more contrasted lines, but also for the He II and the Fe VIII lines (log(T)=5.6). \\

Figure \ref{Fig_Expo} shows that the He II and O IV lines, which form at small temperature, have scaling exponents above two and are larger than the other lines. While the result for the O IV lines is more uncertain since we could not reproduce it using directly the spectrum, the $\alpha$ value for the He II line is more robust; even if it is  larger than the value of 2.16 found for the ESP diode centered around 30.4 nm (see Figure \ref{Fig_304}), this could be explained by the fact that the ESP 304 bandpass includes more emission from "hot" material. It also agrees with the value found for Ly-$\alpha$. Both the He II and H Ly-$\alpha$ lines are strongly optically thick and are coming from the most abundant elements in the Sun. \\
The flare fluences computed for lines forming between about Log(T)=5.5 and Log(T)=6.5 have similar scaling exponents between 1.9 and 2.1. Lines forming at very high temperature, around 10$^7$ K, look to have scaling exponents slightly smaller, around 1.8. These differences are however small and could be explained by uncontroled biases in our analysis; in particular, we are forced to recognize that the lines with the smaller scaling exponent, which are the hotter ones, are also the ones with the larger contrast. \\

This study suggests, but doest not clearly show, that EUV chromospheric lines might have a larger scaling exponent that the coronal emission. We discuss now why it could be so. The fact that different spectral lines might have different scaling exponents can be understood by geometrical consideration:  coronal lines are mostly formed in the hot thermal plasma that is contained in the $V\sim sL$ volume of a loop (with $s$ being the loop footprints area and $L$ the length of the loop), while, during flare, the increase of chromospheric lines relies on the energy received by the loop footpoint and provided at first by the energetic particles (during the impulsive phase) and later on and for longer by conduction. Even if schematic, these different physical processes can account for the fact that chromospheric lines do not scale linearly with coronal ones. This different geometry is also well shown by \cite{Warmuth:2013dp,Warmuth:2013dq} who found, using RHESSI observations of hard X-rays emission for 24 flares, that the thermal source volume of the flares is well correlated with the SXR GOES class while the non-thermal footpoint areas are not.     \\

The fact that $\alpha_{chromosphere} < \alpha_{corona}$ would also mean that when considering smaller flares, in the sense of the X-ray classification, the chromospheric emission becomes more and more important with respect to the coronal one. This can hardly be explained if the chromospheric flare signal depends mostly on the energy provided by conduction from the hot coronal loop. 
The finding of \cite{Warmuth:2013dq} that the non-thermal footpoints area, and therefore one of the parameters governing the energy received by the chromosphere, is not correlated with the soft X-ray class (while thermal volumes are) may explain this behavior. \cite{Longcope:2014th} recently studied flares for which no HXR signatures are observed and where conduction is though to be the only energy driver between the corona and the chromosphere. In his model, the flare energy is deposited at the loop apex and is transported from the corona to the chromosphere by conduction only. Simulating a number of flares with different energies, he found that both the evaporation and condensation velocity scale with the flare energy, but that the condensation velocity has a steeper slope. If we associate the chromospheric emission with the energy brought by condensation, this result is in contradiction with the conclusion suggested in this paper. We also note that there is no evidence in the literature that the flare coronal emission, as observed in SXRs for example, would scale differently depending on the presence or not of HXR signatures of the flare. \\

Finally, if the scaling between flare chromospheric and coronal emissions would hold down to smaller and smaller flares (of class A and B for example) one should see relatively large increase of chromospheric emission for these flares, which is not what is \textit{a priori} observed; one possible explanation would be that the chromospheric contrast, even in solar images, is too small to be detected. Another possibility is that the observed scaling in this study is specific to large (say above the C1 level) flares and that the exponent of the power-law, or, less probably, the shape of the PDF, changes when getting to smaller flares. This will need to be investigated in the future.

\section{Conclusion} 
\label{S-conclu}      

We have investigated the scaling of the flare fluence for different lines formed from slightly above the chromosphere to the corona. Line fluxes already available in the SDO/EVE data have been recomputed and we also identified other lines of interest for this study. In order to deal with small contrasts, we investigated the probability distribution function of the flare fluence for these lines by looking at the scaling of the lines flare fluence with the SXR fluence as observed by GOES in the 0.1-0.8 nm. We found a small tendency for lines forming at very high temperatures to have smaller exponent (about 1.8)  than lines forming between about Log(T)=5.5 and Log(T)=6.5 (around 2). The Lyman-$\alpha$ lines of He II and H seem to have a larger value of the scaling exponent yet, above 2. However, colder lines also have lower contrast and  in despite of our effort, we could not exclude that some biases are responsible for the described behavior. More work is needed to confirm or reject the hypothesis of different scalings for emissions forming at different layers, where the energy is transferred in different ways.

\section{acknowledgements}
 The author acknowledges the SDO/EVE and GOES teams (in particular Janet Machol for the Ly-$\alpha$ data) for providing the data, as well as the organizers and attendants of the "Solar and Stellar Flares" meeting in Prague, 2014, for a very successful and interesting meeting. The author also acknowledge the referee for his comments that significantly helped to improve the paper. Part of this research has been funding by the European Community's Seventh Framework Programme (FP7 2012) under grant agreement no 313188 (SOLID project).

\bibliographystyle{plainnat}

\bibliography{/Users/kretzschmar/BOULOT/BIBLIO/MyBibFromPapers_last.bib}  

\begin{thebibliography}{28}
\providecommand{\natexlab}[1]{#1}
\providecommand{\url}[1]{\texttt{#1}}
\expandafter\ifx\csname urlstyle\endcsname\relax
  \providecommand{\doi}[1]{doi: #1}\else
  \providecommand{\doi}{doi: \begingroup \urlstyle{rm}\Url}\fi

\bibitem[Allred et~al.(2005)Allred, Hawley, Abbett, and
  Carlsson]{2005ApJ...630..573A}
J~C Allred, S~L Hawley, W~P Abbett, and M~Carlsson.
\newblock Radiative hydrodynamic models of the optical and ultraviolet emission
  from solar flares.
\newblock \emph{The Astrophysical Journal}, 630:\penalty0 573--586, 2005.
\newblock \doi{10.1086/431751}.
\newblock URL \url{http://cdsads.u-strasbg.fr/abs/2005ApJ...630..573A}.

\bibitem[{Berlicki}(2007)]{2007ASPC..368..387B}
A.~{Berlicki}.
\newblock {Observations and Modeling of Line Asymmetries in Chromospheric
  Flares}.
\newblock In P.~{Heinzel}, I.~{Dorotovi{\v c}}, and R.~J. {Rutten}, editors,
  \emph{The Physics of Chromospheric Plasmas}, volume 368 of \emph{Astronomical
  Society of the Pacific Conference Series}, pages 387--+, May 2007.

\bibitem[{Berlicki} and {Heinzel}(2004)]{Berlicki:2004fy}
A.~{Berlicki} and P.~{Heinzel}.
\newblock {Soft X-ray heating of the solar chromosphere during the gradual
  phase of two solar flares}.
\newblock \emph{\aap}, 420:\penalty0 319--331, June 2004.
\newblock \doi{10.1051/0004-6361:20035673}.

\bibitem[Crosby et~al.(1993)Crosby, Aschwanden, and Dennis]{Crosby:1993yk}
Norma~B. Crosby, Markus~J. Aschwanden, and Brian~R. Dennis.
\newblock Frequency distributions and correlations of solar x-ray flare
  parameters.
\newblock \emph{Solar Physics}, 143:\penalty0 275--299, February 1993.
\newblock URL \url{http://cdsads.u-strasbg.fr/abs/1993SoPh..143..275C}.

\bibitem[{Dennis}(1985)]{Dennis:1985kq}
B.~R. {Dennis}.
\newblock {Solar hard X-ray bursts}.
\newblock \emph{\solphys}, 100:\penalty0 465--490, October 1985.
\newblock \doi{10.1007/BF00158441}.

\bibitem[{Dere} and {Cook}(1979)]{Dere:1979fj}
K.~P. {Dere} and J.~W. {Cook}.
\newblock {The decay of the 1973 August 9 flare}.
\newblock \emph{\apj}, 229:\penalty0 772--787, April 1979.
\newblock \doi{10.1086/157013}.

\bibitem[Dere et~al.(1997)Dere, Landi, Mason, Fossi, and
  Young]{1997A&AS..125..149D}
K~P Dere, E~Landi, H~E Mason, B~C~Monsignori Fossi, and P~R Young.
\newblock Chianti - an atomic database for emission lines.
\newblock \emph{Astronomy\&Astrophysicss}, 125:\penalty0 149--173, 1997.

\bibitem[{Fletcher} et~al.(2011){Fletcher}, {Dennis}, {Hudson}, {Krucker},
  {Phillips}, {Veronig}, {Battaglia}, {Bone}, {Caspi}, {Chen}, {Gallagher},
  {Grigis}, {Ji}, {Liu}, {Milligan}, and {Temmer}]{Fletcher:2011ys}
L.~{Fletcher}, B.~R. {Dennis}, H.~S. {Hudson}, S.~{Krucker}, K.~{Phillips},
  A.~{Veronig}, M.~{Battaglia}, L.~{Bone}, A.~{Caspi}, Q.~{Chen},
  P.~{Gallagher}, P.~T. {Grigis}, H.~{Ji}, W.~{Liu}, R.~O. {Milligan}, and
  M.~{Temmer}.
\newblock {An Observational Overview of Solar Flares}.
\newblock \emph{\ssr}, 159:\penalty0 19--106, September 2011.
\newblock \doi{10.1007/s11214-010-9701-8}.

\bibitem[{Hannah} et~al.(2011){Hannah}, {Hudson}, {Battaglia}, {Christe},
  {Ka{\v s}parov{\'a}}, {Krucker}, {Kundu}, and {Veronig}]{Hannah:2011vn}
I.~G. {Hannah}, H.~S. {Hudson}, M.~{Battaglia}, S.~{Christe}, J.~{Ka{\v
  s}parov{\'a}}, S.~{Krucker}, M.~R. {Kundu}, and A.~{Veronig}.
\newblock {Microflares and the Statistics of X-ray Flares}.
\newblock \emph{\ssr}, 159:\penalty0 263--300, September 2011.
\newblock \doi{10.1007/s11214-010-9705-4}.

\bibitem[{Heinzel} and {Avrett}(2012)]{Heinzel:2012kx}
P.~{Heinzel} and E.~H. {Avrett}.
\newblock {Optical-to-Radio Continua in Solar Flares}.
\newblock \emph{\solphys}, 277:\penalty0 31--44, March 2012.
\newblock \doi{10.1007/s11207-011-9823-5}.

\bibitem[Hudson(1991)]{Hudson:1991aa}
H~S Hudson.
\newblock Solar flares, microflares, nanoflares, and coronal heating.
\newblock \emph{Solar Physics}, 133:\penalty0 357--369, 1991.

\bibitem[{Hudson}(2011)]{Hudson:2011bh}
H.~S. {Hudson}.
\newblock {Global Properties of Solar Flares}.
\newblock \emph{\ssr}, pages 7--+, January 2011.
\newblock \doi{10.1007/s11214-010-9721-4}.

\bibitem[{Kretzschmar}(2011)]{Kretzschmar:2011lr}
M.~{Kretzschmar}.
\newblock {The Sun as a star: observations of white-light flares}.
\newblock \emph{Astronomy\&Astrophysics}, 530:\penalty0 A84+, June 2011.
\newblock \doi{10.1051/0004-6361/201015930}.

\bibitem[{Kretzschmar} et~al.(2010){Kretzschmar}, {Dudok de Wit}, {Schmutz},
  {Mekaoui}, {Hochedez}, and {Dewitte}]{Kretzschmar:2010lr}
M.~{Kretzschmar}, T.~{Dudok de Wit}, W.~{Schmutz}, S.~{Mekaoui}, {J.-F.}
  {Hochedez}, and S.~{Dewitte}.
\newblock {The effect of flares on total solar irradiance}.
\newblock \emph{Nature Physics}, 6:\penalty0 690--692, September 2010.
\newblock \doi{10.1038/nphys1741}.

\bibitem[Landi et~al.(2006)Landi, Zanna, Young, Dere, Mason, and
  Landini]{2006ApJS..162..261L}
E~Landi, G~Del Zanna, P~R Young, K~P Dere, H~E Mason, and M~Landini.
\newblock Chianti-an atomic database for emission lines. vii. new data for
  x-rays and other improvements.
\newblock \emph{The Astrophysical Journals}, 162:\penalty0 261--280, 2006.
\newblock \doi{10.1086/498148}.

\bibitem[{Longcope}(2014)]{Longcope:2014th}
D.~W. {Longcope}.
\newblock {A Simple Model of Chromospheric Evaporation and Condensation Driven
  Conductively in a Solar Flare}.
\newblock \emph{\apj}, 795:\penalty0 10, November 2014.
\newblock \doi{10.1088/0004-637X/795/1/10}.

\bibitem[{Milligan} et~al.(2012){Milligan}, {Chamberlin}, {Hudson}, {Woods},
  {Mathioudakis}, {Fletcher}, {Kowalski}, and {Keenan}]{Milligan:2012fj}
R.~O. {Milligan}, P.~C. {Chamberlin}, H.~S. {Hudson}, T.~N. {Woods},
  M.~{Mathioudakis}, L.~{Fletcher}, A.~F. {Kowalski}, and F.~P. {Keenan}.
\newblock {Observations of Enhanced Extreme Ultraviolet Continua during an
  X-Class Solar Flare Using SDO/EVE}.
\newblock \emph{\apjl}, 748:\penalty0 L14, March 2012.
\newblock \doi{10.1088/2041-8205/748/1/L14}.

\bibitem[{Milligan} et~al.(2014){Milligan}, {Kerr}, {Dennis}, {Hudson},
  {Fletcher}, {Allred}, {Chamberlin}, {Ireland}, {Mathioudakis}, and
  {Keenan}]{Milligan:2014mz}
R.~O. {Milligan}, G.~S. {Kerr}, B.~R. {Dennis}, H.~S. {Hudson}, L.~{Fletcher},
  J.~C. {Allred}, P.~C. {Chamberlin}, J.~{Ireland}, M.~{Mathioudakis}, and
  F.~P. {Keenan}.
\newblock {The Radiated Energy Budget of Chromospheric Plasma in a Major Solar
  Flare Deduced from Multi-wavelength Observations}.
\newblock \emph{\apj}, 793:\penalty0 70, October 2014.
\newblock \doi{10.1088/0004-637X/793/2/70}.

\bibitem[{Schrijver} et~al.(2012){Schrijver}, {Beer}, {Baltensperger},
  {Cliver}, {G{\"u}del}, {Hudson}, {McCracken}, {Osten}, {Peter}, {Soderblom},
  {Usoskin}, and {Wolff}]{Schrijver:2012rt}
C.~J. {Schrijver}, J.~{Beer}, U.~{Baltensperger}, E.~W. {Cliver},
  M.~{G{\"u}del}, H.~S. {Hudson}, K.~G. {McCracken}, R.~A. {Osten}, T.~{Peter},
  D.~R. {Soderblom}, I.~G. {Usoskin}, and E.~W. {Wolff}.
\newblock {Estimating the frequency of extremely energetic solar events, based
  on solar, stellar, lunar, and terrestrial records}.
\newblock \emph{Journal of Geophysical Research (Space Physics)}, 117:\penalty0
  8103, 2012.
\newblock \doi{10.1029/2012JA017706}.

\bibitem[{Temmer} et~al.(2001){Temmer}, {Veronig}, {Hanslmeier}, {Otruba}, and
  {Messerotti}]{Temmer:2001kq}
M.~{Temmer}, A.~{Veronig}, A.~{Hanslmeier}, W.~{Otruba}, and M.~{Messerotti}.
\newblock {Statistical analysis of solar H{$\alpha$} flares}.
\newblock \emph{\aap}, 375:\penalty0 1049--1061, September 2001.
\newblock \doi{10.1051/0004-6361:20010908}.

\bibitem[{Veronig} et~al.(2002{\natexlab{a}}){Veronig}, {Temmer}, {Hanslmeier},
  {Otruba}, and {Messerotti}]{Veronig:2002qf}
A.~{Veronig}, M.~{Temmer}, A.~{Hanslmeier}, W.~{Otruba}, and M.~{Messerotti}.
\newblock {Temporal aspects and frequency distributions of solar soft X-ray
  flares}.
\newblock \emph{\aap}, 382:\penalty0 1070--1080, February 2002{\natexlab{a}}.
\newblock \doi{10.1051/0004-6361:20011694}.

\bibitem[{Veronig} et~al.(2002{\natexlab{b}}){Veronig}, {Vr{\v s}nak},
  {Dennis}, {Temmer}, {Hanslmeier}, and {Magdaleni{\'c}}]{Veronig:2002zp}
A.~{Veronig}, B.~{Vr{\v s}nak}, B.~R. {Dennis}, M.~{Temmer}, A.~{Hanslmeier},
  and J.~{Magdaleni{\'c}}.
\newblock {Investigation of the Neupert effect in solar flares. I. Statistical
  properties and the evaporation model}.
\newblock \emph{Astronomy\&Astrophysics}, 392:\penalty0 699--712, September
  2002{\natexlab{b}}.
\newblock \doi{10.1051/0004-6361:20020947}.

\bibitem[{Veronig} et~al.(2010){Veronig}, {Ryb{\'a}k}, {G{\"o}m{\"o}ry},
  {Berkebile-Stoiser}, {Temmer}, {Otruba}, {Vr{\v s}nak}, {P{\"o}tzi}, and
  {Baumgartner}]{Veronig:2010uq}
A.~M. {Veronig}, J.~{Ryb{\'a}k}, P.~{G{\"o}m{\"o}ry}, S.~{Berkebile-Stoiser},
  M.~{Temmer}, W.~{Otruba}, B.~{Vr{\v s}nak}, W.~{P{\"o}tzi}, and
  D.~{Baumgartner}.
\newblock {Multiwavelength Imaging and Spectroscopy of Chromospheric
  Evaporation in an M-class Solar Flare}.
\newblock \emph{The Astrophysical Journal}, 719:\penalty0 655--670, August
  2010.
\newblock \doi{10.1088/0004-637X/719/1/655}.

\bibitem[{Viereck} et~al.(2007){Viereck}, {Hanser}, {Wise}, {Guha}, {Jones},
  {McMullin}, {Plunket}, {Strickland}, and {Evans}]{Viereck:2007fk}
R.~{Viereck}, F.~{Hanser}, J.~{Wise}, S.~{Guha}, A.~{Jones}, D.~{McMullin},
  S.~{Plunket}, D.~{Strickland}, and S.~{Evans}.
\newblock {Solar extreme ultraviolet irradiance observations from GOES: design
  characteristics and initial performance}.
\newblock In \emph{Society of Photo-Optical Instrumentation Engineers (SPIE)
  Conference Series}, volume 6689 of \emph{Society of Photo-Optical
  Instrumentation Engineers (SPIE) Conference Series}, page~0, September 2007.
\newblock \doi{10.1117/12.734886}.

\bibitem[{Warmuth} and {Mann}(2013{\natexlab{a}})]{Warmuth:2013dp}
A.~{Warmuth} and G.~{Mann}.
\newblock {Thermal and nonthermal hard X-ray source sizes in solar flares
  obtained from RHESSI observations. I. Observations and evaluation of
  methods}.
\newblock \emph{\aap}, 552:\penalty0 A86, April 2013{\natexlab{a}}.
\newblock \doi{10.1051/0004-6361/201219354}.

\bibitem[{Warmuth} and {Mann}(2013{\natexlab{b}})]{Warmuth:2013dq}
A.~{Warmuth} and G.~{Mann}.
\newblock {Thermal and nonthermal hard X-ray source sizes in solar flares
  obtained from RHESSI observations. II. Scaling relations and temporal
  evolution}.
\newblock \emph{\aap}, 552\penalty0 (A87), 2013{\natexlab{b}}.

\bibitem[Woods et~al.(2006)Woods, Kopp, and Chamberlin]{Woods:2006aa}
T~N Woods, G.~Kopp, and P~C Chamberlin.
\newblock Contributions of the solar ultraviolet irradiance to the total solar
  irradiance during large flares.
\newblock \emph{J. Geophys. Res.}, 111:\penalty0 A10S14, 2006.
\newblock \doi{10.1029/2005JA011507}.

\bibitem[{Woods} et~al.(2012){Woods}, {Eparvier}, {Hock}, {Jones}, {Woodraska},
  {Judge}, {Didkovsky}, {Lean}, {Mariska}, {Warren}, {McMullin}, {Chamberlin},
  {Berthiaume}, {Bailey}, {Fuller-Rowell}, {Sojka}, {Tobiska}, and
  {Viereck}]{Woods:2010zr}
T.~N. {Woods}, F.~G. {Eparvier}, R.~{Hock}, A.~R. {Jones}, D.~{Woodraska},
  D.~{Judge}, L.~{Didkovsky}, J.~{Lean}, J.~{Mariska}, H.~{Warren},
  D.~{McMullin}, P.~{Chamberlin}, G.~{Berthiaume}, S.~{Bailey},
  T.~{Fuller-Rowell}, J.~{Sojka}, W.~K. {Tobiska}, and R.~{Viereck}.
\newblock {Extreme Ultraviolet Variability Experiment (EVE) on the Solar
  Dynamics Observatory (SDO): Overview of Science Objectives, Instrument
  Design, Data Products, and Model Developments}.
\newblock \emph{Solar Physics}, 275\penalty0 (1-2):\penalty0 115--143, January
  2012.
\newblock \doi{10.1007/s11207-009-9487-6}.

\end{thebibliography}



\end{document}